\documentclass[aps,showpacs,twocolumn,superscriptaddress,prb,preprintnumbers,nofootinbib]{revtex4}
\usepackage{amsmath,amssymb,bm,graphicx,color}

\begin{document}

\title{Order and excitations in large-$S$ kagom\'{e}-lattice antiferromagnets}

\author{A. L. Chernyshev}
\affiliation{Department of Physics and Astronomy, University of California, Irvine, California
92697, USA}
\author{M. E. Zhitomirsky}
\affiliation{Service de Physique Statistique, Magn\'etisme et Supraconductivit\'e,
UMR-E9001 CEA-INAC/UJF, 17 rue des Martyrs, 38054 Grenoble Cedex 9, France}
\date{\today}
\begin{abstract} 
We systematically investigate the ground-state and the spectral properties of antiferromagnets 
on a kagom\'{e} lattice with several common types of the planar anisotropy:  $XXZ$,  single-ion, and  
out-of-plane Dzyaloshinskii-Moriya. Our main focus is on the role of nonlinear, anharmonic terms, which 
are responsible for the quantum order-by-disorder effect and for the corresponding selection 
of the ground-state spin structure in many of these models. The $XXZ$ and the single-ion anisotropy models 
exhibit a quantum phase transition between the ${\bf q}\!=\!0$ and the $\sqrt{3}\times\!\sqrt{3}$ states
as a function of the anisotropy parameter, offering a rare example of the quantum order-by-disorder fluctuations
favoring a ground state which is different from the one selected by  thermal fluctuations.
The nonlinear terms are also shown to be crucial for a very strong near-resonant decay phenomenon 
leading to the quasiparticle breakdown in the  kagom\'{e}-lattice antiferromagnets whose spectra 
are featuring flat or weakly dispersive modes.
The effect is shown to persist even in the limit of large spin values  and should be common to  other frustrated 
magnets with flat branches of excitations. Model calculations of the spectrum of the $S=5/2$ Fe-jarosite 
with Dzyaloshinskii-Moriya anisotropy
provide a convincing and detailed characterization of the proposed scenario.
\end{abstract}
\maketitle

\section{Introduction}

Kagom\'{e}-lattice antifferomagnets are iconic in the field of frustrated magnets, comprising a 
number of model systems whose classical ground states are massively degenerate, 
giving rise to an extreme sensitivity to subtle symmetry breaking effects,\cite{Huse92,Chalker92} 
to fractional magnetization plateaus, \cite{MZH02,Capponi13}
to a strongly amplified role of secondary interactions,\cite{Harris92} and to an emergent hierarchy of energy scales
in their dynamics.\cite{Taillefumier14} A crucial role of the non-linear, anharmonic terms in the so-called
order-by-disorder ground-state selection  by thermal\cite{Reimers93} or quantum fluctuations\cite{Chubukov92} in 
the kagom\'{e}-lattice antifferomagnets has been recognized 
for some time. Recently, an accurate, systematic treatment of the 
quantum order-by-disorder effect in the anisotropic versions of the model has 
received a significant development.\cite{ChZh14,Gotze15}
However, much less is known about the role of such terms in the excitation spectra of 
frustrated magnets and only recently a rather dramatic picture has begun to emerge. \cite{Ch15}

Usually, the  nonlinear  terms in antiferromagnets are 
necessary to describe interactions of magnons and, while their role is, generally, more significant 
when frustration is present,\cite{RMP13,triPRB09} they still lead to effects that are relatively minor in the large-$S$ limit,
i.e., constitute a $1/S$ contribution compared with the classical energy scale $JS$. 
However, in a wide class of highly-frustrated systems,  including  the kagom\'e-lattice antiferromagnets, the
non-linear anharmonic terms are responsible for the phenomena that are much more dramatic. 

First effect concerns   systems in which contenders for the ground state form a highly degenerate, extensive 
manifold of states, and neither the classical energetics nor the harmonic fluctuations
are able to select  a unique ground state from that manifold.\cite{Chalker92,Chubukov92,Harris92} 
In that case, the selection role is passed onto the nonlinear terms 
providing a variant of the quantum order by disorder effect.
Such is the case of the three models considered in this work, Heisenberg, $XXZ$,\cite{ChZh14}  
and single-ion anisotropy models, 
while in the case of the Dzyaloshinskii-Moriya (DM) anisotropy, a unique state is selected already on the 
classical level.\cite{Cepas08}

The second effect is less-studied, but is equally striking. 
We demonstrate, that the nonlinear terms  lead  to spectacularly strong quantum effects 
in the dynamical response of the flat-band frustrated magnets, even in the ones that are assumed nearly classical.\cite{Ch15}
The resultant spectral features invoke parallels with the quasiparticle breakdown signatures in quantum spin-
and Bose-liquids,\cite{Zaliznyak,Zheludev}
which exhibit  termination points and  broad continua where single-particle excitations are no longer well-defined. 
In the present case, the origin of such features is in the near-resonant decay into   pairs of the flat modes, 
facilitated by the nonlinear couplings. The effect is strongly amplified by the density of states 
of the flat modes and can be shown to persist even in the large-$S$ limit, defying the usual $1/S$ suppression 
trend and challenging  a conventional wisdom that such drastic 
phenomena can only occur in an inherently quantum  system.

In the present study, we expand the analysis of our previous works\cite{ChZh14,Ch15} 
and offer an expos\'{e} of the  $1/S$ formalism for several common anisotropic extensions
of the  nearest-neighbor  Heisenberg model on the kagom\'{e} lattice that are also relevant to real materials.
The presented  approach to the nonlinear spin-wave theory can be applicable to the other, more complicated 
forms of the kagom\'{e}-lattice Hamiltonians as well as to a broader class of frustrated spin systems on the
non-Bravais lattices. We also provide a useful extension of our approach to the perturbative treatment
of small perturbations to the main Hamiltonian, such as the next-nearest superexchanges $J_2$. 

For the ground-state   consideration, we 
demonstrate a quantum phase transition between the ${\bf q}\!=\!0$ and the $\sqrt{3}\times\!\sqrt{3}$ states
as a function of  anisotropy parameter  in two models: $XXZ$ model, studied previously,\cite{ChZh14} and the single-ion 
anisotropy model. While the effect is similar in both models, the energy scale associated with it is 
shown to be different, in agreement with the understanding that the degeneracy-lifting interaction is
associated with a high-order topologically-nontrivial spin-flip processes, which are different in the two models.
Nevertheless, both model cases  present a rare example  of the ground-state selection that is different 
from the choice of the  thermal fluctuations, which 
favor the $\sqrt{3}\times\!\sqrt{3}$  structure for any value of the anisotropy parameter. 
\cite{Harris92,Reimers93,Henley09,Huse92,Korshunov02,Rzchowski97,ChZh14,Gotze15} 

For the spectral properties of the kagom\'{e}-lattice antiferromagnets, we offer a detailed consideration
of the decay-induced effects in the DM anisotropy model with $J_2$, the model that closely 
describes the $S=5/2$ Fe-jarosite.\cite{Matan06,Yildirim06} We also present a general analysis 
of the near-resonant decays in the flat-band frustrated antiferromagnets, which suggests that
the dramatic modifications in the spectrum due to this phenomenon must persist in the large-$S$ limit.
While the core of our presentation is aimed at a common and realistic extension of the nearest-neighbor Heisenberg 
kagom\'{e}-lattice antiferromagnet,
we argue that the spectacularly strong quantum effect of   the quasiparticle breakdown in an almost classical system
should be applicable to a variety of other flat-band frustrated spin systems.
\cite{Petrenko,Matan14,Balents_honeycomb,Georgii15} 
We also remark on a useful ${\bf q}$-dependence of the dynamical structure factor, which  
is characteristic to the non-Bravais lattices and allows to select spectral
contributions of specific  branches  in the portions of  the ${\bf q}$-space.

The paper is organized as follows. In Sec.~\ref{Sec:linear} we present a 
 consideration of the harmonic theory for several common anisotropic models 
on the kagom\'{e} lattice, explicate    details of the diagonalization procedure, and
show results of a representative calculation  within the harmonic approximation. 
In Sec.~\ref{Sec:nonlinear}, anharmonic terms of the models are derived and the results of the ground-state 
selection calculations are presented. The spectral properties of the kagom\'{e}-lattice antiferromagnets are  also
given a detailed exposition. Technical aspects of the derivation of the quartic terms are given in
Appendix~\ref{AppA}.

\section{Linear spin-wave theory}
\label{Sec:linear}

To set the stage, we   provide the linear spin-wave   consideration of the 
isotropic nearest-neighbor antiferromagnetic Heisenberg model on the kagom\'e lattice  
following the approach of Ref.~\onlinecite{Harris92}.
We continue with    various extensions of the model, which are either relevant to real materials or 
allow to explore the role of quantum effects in a wider parameter space. These extensions 
include the anisotropic $XXZ$ model, models with the single-ion and the out-of-plane 
Dzyaloshinskii-Moria anisotropies, and additional further-neighbor exchange terms.
Subsequently, the important steps of the diagonalization procedure that will be essential for the non-linear 
terms considered in Sec.~\ref{Sec:nonlinear} are exposed.
In the end of this Section, results of the calculations of the on-site magnetization within 
the  linear spin-wave theory for the $XXZ$ model are presented as an example.   

\subsection{Nearest-neighbor Heisenberg model}

Within the spin-wave treatment, nearest-neighbor Heisenberg antiferromagnet 
on the kagom\'e lattice with the Hamiltonian 
\begin{equation}
\hat{\cal H} = J \sum_{\langle ij\rangle} {\bf S}_i\cdot {\bf S}_j \ ,
\label{Has}
\end{equation}
is assumed to be in an ordered state with  spins forming
a coplanar 120$^\circ$ structure.  Summation is over bonds and $i,j$ are the sites of the  lattice.
Aligning the $z$-axis on each site along  the direction of the ordered moment and directing the 
$y$-axes  out of the ordering plane transforms the Hamiltonian (\ref{Has}) to a local spin basis 
\begin{eqnarray}
\hat{\cal H} &=& J \sum_{\langle ij\rangle} \Bigl(  S_i^yS_j^y +
\cos \theta_{ij} \left(S_i^xS_j^x + S_i^z S_j^z\right) \nonumber\\
\label{Hs}
&&+\sin \theta_{ij} \left(S_i^zS_j^x - S_i^x S_j^z\right) \Bigr)
= J \sum_{\langle ij\rangle} {\bf S}_i \otimes{\bf S}_j
\ ,
\end{eqnarray}
where $\theta_{ij}= \theta_i - \theta_j$ is an angle between two neighboring spins and we have
introduced ``matrix'' product $\otimes$ of spins  with the matrix
\begin{equation}
\label{matrix}
\otimes=\left(\begin{array}{ccc}
\cos \theta_{ij}   &  0 & -\sin \theta_{ij} \\
0 &  1   & 0 \\
\sin \theta_{ij} &  0 &  \cos \theta_{ij}
\end{array}\right) \, 
\end{equation}
as a shorthand notation.
\begin{figure}[tb]
\includegraphics[width=0.99\columnwidth]{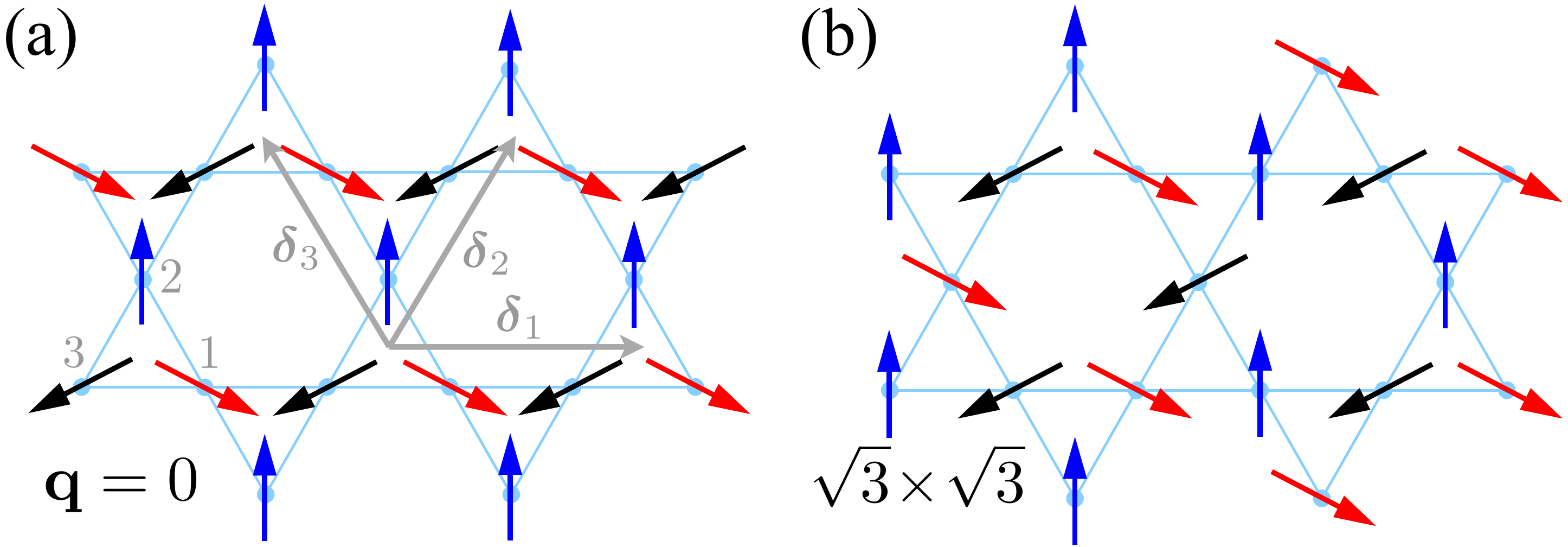}
\caption{(Color online)\   
 (a) ${\bf q}\!=\!0$  and (b)  $\sqrt{3}\times\!\sqrt{3}$ spin configurations. In (a), primitive vectors 
 of the kagom\'e lattice and numbering of sites within the unit cell are shown.
}
\label{basis}
\end{figure}

We choose the unit cell of the kagom\'e lattice as an up-triangle, see Fig.~\ref{basis},
and the primitive vectors of the corresponding triangular Bravais lattice as 
\begin{eqnarray}
\bm{\delta}_1 = (1,0)\, , \
\bm{\delta}_2 = \biggl(\frac{1}{2}, \frac{\sqrt{3}}{2}\biggr)\, , \
\bm{\delta}_3 = \bm{\delta}_2-\bm{\delta}_1\, ,
\end{eqnarray}
all in units of $2a$ where $a$ is the interatomic distance.
The  atomic coordinates within the unit cell are  $\bm{\rho}_1 = 0$, $\bm{\rho}_2 = \frac{1}{2}\bm{\delta}_3$, and
$\bm{\rho}_3 = -\frac{1}{2}\bm{\delta}_1$.

Then, changing summation over bonds to summation over unit cells and the atomic index,
$i\rightarrow\{\alpha,\ell\}$, with $\alpha=1$--3 enumerating atoms within the unit cell, 
the Hamiltonian \eqref{Hs} becomes
\begin{eqnarray}
\label{Hls}
\hat{\cal H} &=& J \sum_\ell {\bf S}_{1,\ell}\otimes\left({\bf S}_{2,\ell} + {\bf S}_{2,\ell-3}\right)  \\
&&+ {\bf S}_{1,\ell}\otimes\left({\bf S}_{3,\ell} + {\bf S}_{3,\ell+1}\right) +
{\bf S}_{2,\ell}\otimes\left({\bf S}_{3,\ell} + {\bf S}_{3,\ell+2}\right) \, ,\nonumber
\end{eqnarray}
where the product ${\bf S}_{\alpha,\ell}\otimes {\bf S}_{\alpha',\ell'}$ is according to (\ref{Hs}),
and $\ell\pm n\equiv{\bf R}_\ell \pm\bm{\delta}_n$ with the coordinate of the unit cell
${\bf R}_\ell= m_1\bm{\delta}_1 + m_2\bm{\delta}_2$.

\subsection{Harmonic spin-wave approximation}

It should be noted that although a coplanar state with spins on each triangle 
in a 120$^\circ$ structure  minimizes the classical energy of (\ref{Has}), 
such a state  is not unique and the manifold of them is extensive.\cite{Baxter70} 
However, it is also clear from the  Hamiltonian in the local basis (\ref{Hs}) that the  linear
spin-wave theory is the same for \emph{any} state from this   degenerate 
manifold,\cite{Chalker92} because  $\cos \theta_{ij}=-1/2$ for any pair of spins in such state  
and $S^x S^z$ terms do not contribute  to the harmonic order of the $1/S$ expansion.

Thus,    we introduce  Holstein-Primakoff
representation for spin operators  
\begin{eqnarray}
S_{\alpha,\ell}^z=S- a^\dagger_{\alpha,\ell}a^{\phantom{\dag}}_{\alpha,\ell}\ , \ \ 
S_{\alpha,\ell}^-=a_{\alpha,\ell}^\dag\sqrt{2S- a^\dagger_{\alpha,\ell}a^{\phantom{\dag}}_{\alpha,\ell}} 
\label{HP}
\end{eqnarray} 
into \eqref{Hls}  and, keeping only quadratic  terms, obtain a harmonic Hamiltonian
for  three species of bosons
\begin{eqnarray}
\hat{\cal H}_2 & = & 2JS \sum_\ell \biggl\{\Bigl[a_{1,\ell}^\dagger a^{\phantom{\dag}}_{1,\ell} +
a_{2,\ell}^\dagger a^{\phantom{\dag}}_{2,\ell} + a_{3,\ell}^\dagger a^{\phantom{\dag}}_{3,\ell}\Bigr] \nonumber\\
&+&
\frac{1}{8}
\Bigl[a_{1,\ell}^\dagger\bigl(a^{\phantom{\dag}}_{2,\ell}+a^{\phantom{\dag}}_{2,\ell-3}\bigr) 
+ a_{1,\ell}^\dagger\bigl(a^{\phantom{\dag}}_{3,\ell} +a^{\phantom{\dag}}_{3,\ell+1}\bigr)\nonumber\\
&&\phantom{\frac{1}{8}\Bigl[}
+\,  a_{2,\ell}^\dagger\bigl(a^{\phantom{\dag}}_{3,\ell}+a^{\phantom{\dag}}_{3,\ell+2}\bigr) + \textrm{h.c.}\Bigr]
\label{H2R} \\
&-&
 \frac{3}{8}
\Bigl[a^{\phantom{\dag}}_{1,\ell}\bigl(a^{\phantom{\dag}}_{2,\ell}+a^{\phantom{\dag}}_{2,\ell-3}\bigr) + 
a^{\phantom{\dag}}_{1,\ell}\bigl(a^{\phantom{\dag}}_{3,\ell}+a^{\phantom{\dag}}_{3,\ell+1}\bigr)\nonumber\\
&& \phantom{\frac{3}{8}\Bigl[}
+\,  a^{\phantom{\dag}}_{2,\ell}\bigl(a^{\phantom{\dag}}_{3,\ell}+a^{\phantom{\dag}}_{3,\ell+2}\bigr) + \textrm{h.c.}\Bigr]
\biggr\} \ .
\nonumber
\end{eqnarray}
Performing the Fourier transformation according to
\begin{equation}
a^{\phantom{\dag}}_{\alpha,\ell} = \frac{1}{\sqrt{N}} \sum_{\bf k}  
a^{\phantom{\dag}}_{\alpha,\bf k} \, e^{i{\bf k}{\bf r}_{\alpha,\ell}},
\end{equation}
where ${\bf r}_{\alpha,\ell}\! =\! {\bf R}_\ell\! +\! \bm{\rho}_\alpha$ and $N$ is the number of unit cells,
we obtain the linear spin-wave theory Hamiltonian
\begin{eqnarray}
\hat{\cal H}_2 &=& 2 JS \sum_{{\bf k},\alpha\beta}\biggl\{\Bigl[\delta_{\alpha \beta}+
\frac{1}{4}\,
\Lambda^{\alpha\beta}_{\bf k}\Bigr] a_{\alpha,\bf k}^\dagger a^{\phantom{\dag}}_{\beta,\bf k} \nonumber\\
&&\phantom{2J\sum_{{\bf k},\alpha\beta}\biggl\{}- \frac{3}{8}\,\Lambda^{\alpha\beta}_{\bf k}\bigl(
a_{\alpha,\bf k}^\dagger a^\dagger_{\beta,-\bf k}+\textrm{h.c.}\bigr)\biggr\},\label{H2F}
\end{eqnarray}
with  the matrix
\begin{equation}
\label{Mk}
\hat{\bm\Lambda}_{\bf k} =\left(\begin{array}{ccc}
0   &  c_3 & c_1 \\
c_3 &  0   & c_2 \\
c_1 &  c_2 &  0
\end{array}\right) \, ,
\end{equation}
and  shorthand notations $c_n = \cos(q_n)$ with $q_n\!=\!{\bf k}\cdot\bm{\delta}_n/2$.

One can rewrite this Hamiltonian as
\begin{equation}
\label{Hmatrix}
\hat{\cal H}_2 =  \sum_{\mathbf{k}>0}
\hat{X}^\dagger_{\bf k} \hat{\bf H}_{\bf k} \hat{X}_{\bf k} - 3JS\, ,
\end{equation}
with  the vector operator
\begin{equation}
\label{Xvector}
\hat{X}^\dagger_{\bf k} \!=\!
\bigl(a_{1,\bf k}^\dagger, a_{2,\bf k}^\dagger, a_{3,\bf k}^\dagger,
a^{\phantom{\dag}}_{1,-\bf k},a^{\phantom{\dag}}_{2,-\bf k},a^{\phantom{\dag}}_{3,-\bf k}\bigr)
\end{equation}
and the $6\!\times\! 6$ matrix $\hat{\bf H}_{\bf k}$
\begin{equation}
\hat{\bf H}_{\bf k} = 2J S\left(\begin{array}{cc}
\hat{\bf A}_{\bf k} &  -\hat{\bf B}_{\bf k}  \\
-\hat{\bf B}_{\bf k}  &    \hat{\bf A}_{\bf k}
\end{array}\right)\, , \label{Hmatrix1}
\end{equation}
where
\begin{equation}
\hat{\bf A}_{\bf k} =\hat{\bf I}+\frac{1}{4} \, \hat{\bm\Lambda}_{\bf k}, \ \
\hat{\bf B}_{\bf k} =\frac{3}{4} \, \hat{\bm\Lambda}_{\bf k} \, ,
\label{AB}
\end{equation}
and $\hat{\bf I}$ is the identity matrix.

For a moment, we will confine ourselves to the eigenvalue problem of $\hat{\cal H}_2$.
Because of an obvious commutativity of the matrices $\hat{\bf A}_{\bf k}$ and $\hat{\bf B}_{\bf k}$, 
 the eigenvalues of $\hat{\bf H}_{\bf k}$
are straightforwardly related to their eigenvalues, and, in turn, are
determined by the eigenvalues $\lambda_{\nu,\bf k}$ of the matrix $\hat{\bm\Lambda}_{\bf k}$, so that
the spin-wave excitation energies are
\begin{equation}
\varepsilon_{\nu,\bf k} = 2JS\sqrt{A_{\nu,\bf k}^2-B_{\nu,\bf k}^2}
=  2JS \omega_{\nu,\bf k} \ ,
\label{Ek}
\end{equation}
with the frequencies $\omega_{\nu,\bf k}=\sqrt{(1-\lambda_{\nu,\bf k}/2)(1+ \lambda_{\nu,\bf k})}$ and
\begin{equation}
A_{\nu,\bf k} =1+\frac{1}{4} \, \lambda_{\nu,\bf k}, \
B_{\nu,\bf k} =\frac{3}{4} \, \lambda_{\nu,\bf k} \, .
\label{AB1}
\end{equation}
Thus, the problem of the diagonalization of $\hat{\cal H}_2$ in (\ref{H2F}) is reduced to the
eigenvalue problem of $\hat{\bm\Lambda}_{\bf k}$ in (\ref{Mk}).
From the characteristic equation for  $\hat{\bm\Lambda}_{\bf k}$  one finds
\begin{equation}
|\hat{\bm\Lambda}_{\bf k} -\lambda | =
\left(\lambda +1\right) \left(\lambda^2 - \lambda - 2\gamma_{\bf k}\right) = 0 \ ,
\label{lambda}
\end{equation}
where $\gamma_{\bf k} \equiv c_1 c_2 c_3$ is introduced and factorization is performed with the help
of a useful identity
\begin{equation}
c_1^2 + c_2^2 + c_3^2 = 1 + 2 c_1 c_2 c_3\,, 
\nonumber
\end{equation}
which holds once the cosine arguments satisfy $q_2=q_1+q_3$.
Thus, the $\lambda_{\nu,\bf k}$ eigenvalues are
\begin{equation}
\lambda_1 = -1 \ , \quad \lambda_{2(3),{\bf k}} = \frac{1}{2}\,\left(1 \pm \sqrt{1 + 8\gamma_{\bf k}}\right) \, .
\label{lambda123}
\end{equation}
Of the resultant spin-wave excitations one is completely dispersionless and has zero energy, referred to as the ``flat mode,''
and two  are ``normal,'' i.e.  dispersive modes, which are degenerate  in the Heisenberg limit
\begin{equation}
\label{w1}
\varepsilon_{1,\bf k}  = 0, \ \ \ \varepsilon_{2(3),\bf k}  = 2JS \sqrt{1-\gamma_{\bf k}} \ .
\end{equation}
The nature of the flat mode has been discussed previously.\cite{Chubukov92,Harris92,Chalker92}
Generally, such modes owe their origin to both the topological structure of the underlying 
lattices that facilitate spin frustration and the insufficient constraint on the manifold of spin configurations.
Physically, they correspond to the localized, alternating out-of-plane fluctuations of spins around 
elementary hexagons,\cite{Chalker92,Harris92} which do not experience a 
restoring force in the harmonic order in the Heisenberg limit, hence their energy is zero. In the following,
various anisotropies lift the energy of such a flat mode, but preserve its flatness.

\subsection{$XXZ$ model}
\label{XXZ}

Next, we consider an extension of the Heisenberg model on the kagom\'e lattice 
to the  $XXZ$ model with anisotropy of the  easy-plane type, $0\leq\Delta \leq 1$.
The original motivation for this extension, see Ref.~\onlinecite{ChZh14}, was 
that the degeneracy among the $120^\circ$ coplanar states in this model 
remains the same as in the Heisenberg model. This has allowed us to extend the parameter space 
and to study the effect of quantum fluctuations in the ground-state selection 
without lifting degeneracy of the classical ground-state manifold, see Sec.~\ref{Sec:nonlinear} for more detail.\cite{ChZh14}

In the case of the $XXZ$ model, the plane for the coplanar 120$^\circ$ structure is chosen by the
anisotropy. In the local spin basis of (\ref{Hs}) with $y$-axis  
out of the ordering plane, the $XXZ$ addition to the Heisenberg model  reads as  
\begin{equation}
\delta\hat{\cal H} = J(\Delta-1) \sum_{\langle ij\rangle} S^y_i \, S^y_j \, .
\label{Hsxxz}
\end{equation}
A now straightforward spin-wave algebra of (\ref{Hsxxz}) leaves  the  structure of the harmonic 
Hamiltonian in (\ref{H2F}) intact,  yielding corrections to the  
 Hamiltonian matrix in  (\ref{Hmatrix1})   
\begin{eqnarray}
\delta\hat{\bf A}_{\bf k} = \delta\hat{\bf B}_{\bf k} =
\frac{\left(\Delta -1\right)}{2}\,\hat{\bm\Lambda}_{\bf k}\, .
\label{ABxxz}
\end{eqnarray}
The spin-wave energies for the $XXZ$ model in  (\ref{Ek})
are now with $\omega_{\nu,\bf k}=\sqrt{(1-\lambda_{\nu,\bf k}/2)(1+\Delta\lambda_{\nu,\bf k})}$ and give
\begin{equation}
\label{w1xxz}
\varepsilon_{1,\bf k}  = 2JS \sqrt{3(1-\Delta)/2} \, ,
\end{equation}
for  the  flat mode, which is now at a finite energy,   and 
\begin{equation}
\label{w23xxz}
\varepsilon_{2(3),\bf k}  = 2JS \sqrt{1-\Delta\gamma_{\bf k} -(1-\Delta)
\bigl(1 \pm \sqrt{1 + 8\gamma_{\bf k}}\,\bigr)/4}\, .\nonumber
\end{equation}
for the  dispersive modes.

\subsection{Single-ion anisotropy}

Instead of the $XXZ$ correction \eqref{Hsxxz},
an alternative way of generating easy-plane anisotropy is to add a positive single-ion term 
\begin{equation}
\delta\hat{\cal H} =D\sum_i \big(S_i^y\big)^2\ ,
\label{HD}
\end{equation}
where $y$ is the out-of-plane axis in the local spin basis of (\ref{Hs}) as before.
This term, in a complete similarity to the $XXZ$ case, gives zero contribution to the classical energy, 
does not affect cubic anharmonicity, and does not contribute  to 
the degeneracy lifting of the 120$^\circ$ manifold through the quartic terms.\cite{ChZh14}  Its contribution to the
harmonic Hamiltonian (\ref{Hmatrix1}) is also simple
\begin{equation}
\delta\hat{\bf A}_{\bf k} = \delta\hat{\bf B}_{\bf k} =\frac{d}{2}\, \hat{\bf I}  \, ,
\label{ABsi}
\end{equation}
where $d\!=\!D/J$ and $\hat{\bf I}$ is the identity matrix.
Thus, once again, the eigenvalue problem of $\hat{\cal H}_{2}$ reduces to the eigenvalue problem 
of $\hat{\bm\Lambda}_{\bf k}$, resulting in the spin-wave frequencies 
$\omega_{\nu,\bf k}=\sqrt{\left(1-\lambda_{\nu,\bf k}/2\right)\left(1+d+\lambda_{\nu,\bf k}\right)}$,
and the spin-wave energy of the  flat mode   
\begin{equation}
\label{w1si}
\varepsilon_{1,\bf k}  = 2JS \sqrt{3d/2} \, ,
\end{equation}
while energies of the dispersive modes are
\begin{equation}
\label{w23si}
\varepsilon_{2(3),\bf k}  = 2JS \sqrt{1+d-\gamma_{\bf k} -d
\bigl(1 \pm \sqrt{1 + 8\gamma_{\bf k}}\,\bigr)/4}\, ,
\end{equation}
which should be compared to the $XXZ$ results above.

There is a high degree of similarity of the   single-ion anisotropy model (\ref{HD}) and its results to the $XXZ$ case, 
the most important being no degeneracy lifting among the 
120$^\circ$ manifold of classical states at the harmonic level of approximation. However,
there is also an important difference. If analyzed in real space  in the local basis \eqref{Hs}, 
the structure of the spin-flip terms is different in the two models. In particular, the single-ion term \eqref{HD}
creates spin flips that are purely local while the spin-flip hopping 
and other amplitudes  responsible for degeneracy-lifting remain  independent of anisotropy $D$. 
Therefore,  from the point of view of the real-space perturbation theory, described in Ref.~\onlinecite{ChZh14},
the minimal order in which an effective degeneracy-lifting interaction is generated is different in the two models.\cite{Zh_RSunpub}
We will address this difference  in Section~\ref{Sec:nonlinear} in more detail.

\subsection{Dzyaloshinskii-Moria interaction}

Important correction to the Heisenberg spin model (\ref{Has}) that 
commonly occurs in real magnetic materials with the kagom\'{e} structure is the 
anisotropic Dzyaloshinskii-Moriya (DM) interaction.

\begin{figure}[t]
\includegraphics[width=0.65\columnwidth]{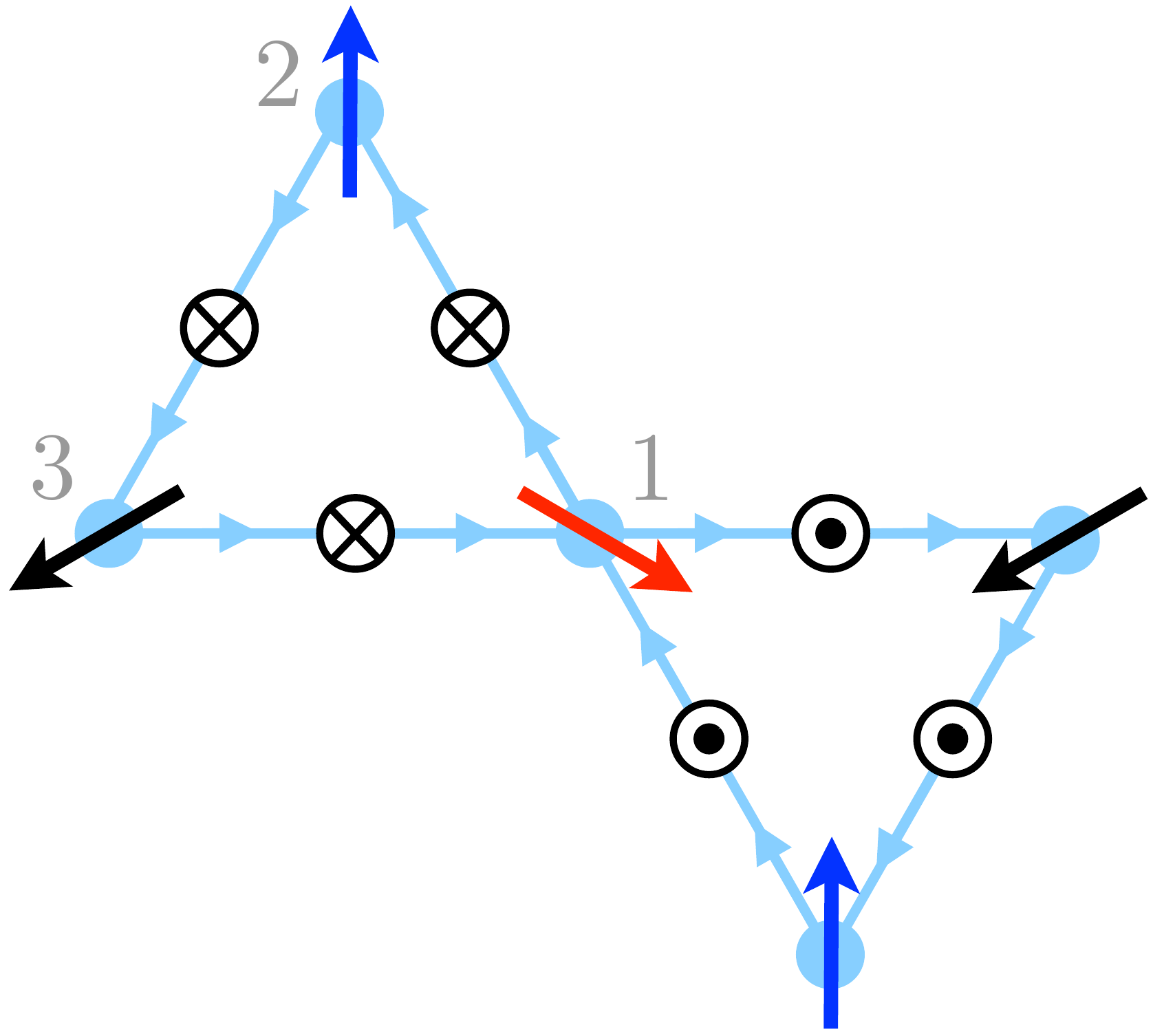}
\caption{(Color online)
Directions of the out-of-plane DM vectors. Arrows on the bonds show the ordering of 
the ${\bf S}_i$ and ${\bf S}_j$ operators in the vector-product in \eqref{HDM}.
}
\label{Fig:DM}
\end{figure}

The out-of-plane DM term, see Fig.~\ref{Fig:DM},
is the main perturbation to the Heisenberg Hamiltonian of the $S =5/2$  kagom\'e-lattice 
antiferromagnet Fe-jarosite,\cite{Matan06,Yildirim06}  $S=1/2$ materials 
herbertsmithite\cite{Zorko08} and vesignieite,\cite{Yoshida12} and other systems.\cite{francisite15}
This anisotropy differs
from the ones considered above  in a number of aspects, but allows for a very similar analytical treatment, 
mainly because of its nearest-neighbor nature.

The antisymmetric Dzyaloshinskii-Moriya interaction is generally written as
\begin{equation}
\delta\hat{\cal H}_{DM} = \sum_{\langle ij\rangle}  {\bf D}_{ij}\cdot ({\bf S}_i \times{\bf S}_j) \, .
\label{HDM}
\end{equation}
In order to determine the DM vectors ${\bf D}_{ij}$ one has to specify the order of sites 
in the vector product, which may be represented by the bond direction from the first to the second spin 
in each pair. A convenient choice consists of ordering spins uniformly along the chains, see Fig.~\ref{Fig:DM}.
Then, the symmetry analysis yields a unique pattern of the DM vectors orthogonal to 
the kagom\'e plane as shown in Fig.~\ref{Fig:DM}. \cite{Elhajal02,Cepas08,Sachdev10}
Note, that the in-plane components of the DM vectors are strictly forbidden once the the kagom\'e plane coincides
with a mirror crystal plane. \cite{Elhajal02} 

Apparent alternation of the DM vectors between up triangles with ${\bf D}_{ij}\!=\!(0,0,-D_z)$ and  
down triangles with ${\bf D}_{ij}\!=\!(0,0,D_z)$ is partly fictitious, because it is a consequence
of the chosen bond gauge, in which
the two types of triangles are circumvented oppositely, Fig.~\ref{Fig:DM}.
The DM interactions on the two triangles favor the same sense of spin rotation or chirality.
Hence, for a given sign of $D_z$, the DM term \eqref{HDM} selects one of the two ${\bf q}\!=\!0$ structures
with positive ($D_z\!>\!0$) or negative ($D_z\!<\!0$) chiralities, yielding  energy gain 
$E_{\rm cl}\!=\!-\sqrt{3} |D_z|S^2$ per site. On the other hand, for the $\sqrt{3}\times\!\sqrt{3}$ state
contributions from up and down triangles come with opposite sign and cancel out. 
In the following, we assume $D_z\!>\!0$ and, consequently, perform the spin-wave expansion around the ${\bf q}\!=\!0$ 
state with positive chiralities, see Fig.~\ref{Fig:DM}. It is also straightforward to see that
the DM term does not induce additional canting and preserves the 120$^\circ$ magnetic structure in the 
$x$--$y$ plane.

Consider the DM term (\ref{HDM}) on the $(1,2)$ bond in Fig.~\ref{Fig:DM}
written in the rotating local spin basis (\ref{Hs})
\begin{eqnarray}
\hat{\cal H}^{(1,2)} &=& D_z\sin\theta_{12}\left(S_2^z S_1^z+S_2^x S_1^x\right)\nonumber \\
&&+D_z\cos\theta_{12}\left(S_2^z S_1^x-S_2^x S_1^z\right)\, ,
\label{HDM12local}
\end{eqnarray}
where $\theta_{ij}\!=\!\theta_i\!-\!\theta_j$, and the first term contributes to the classical, harmonic,
and quartic orders of the $1/S$ expansion, while 
the second contributes in the cubic order.

\begin{figure}[tb]
\includegraphics[width=0.99\columnwidth]{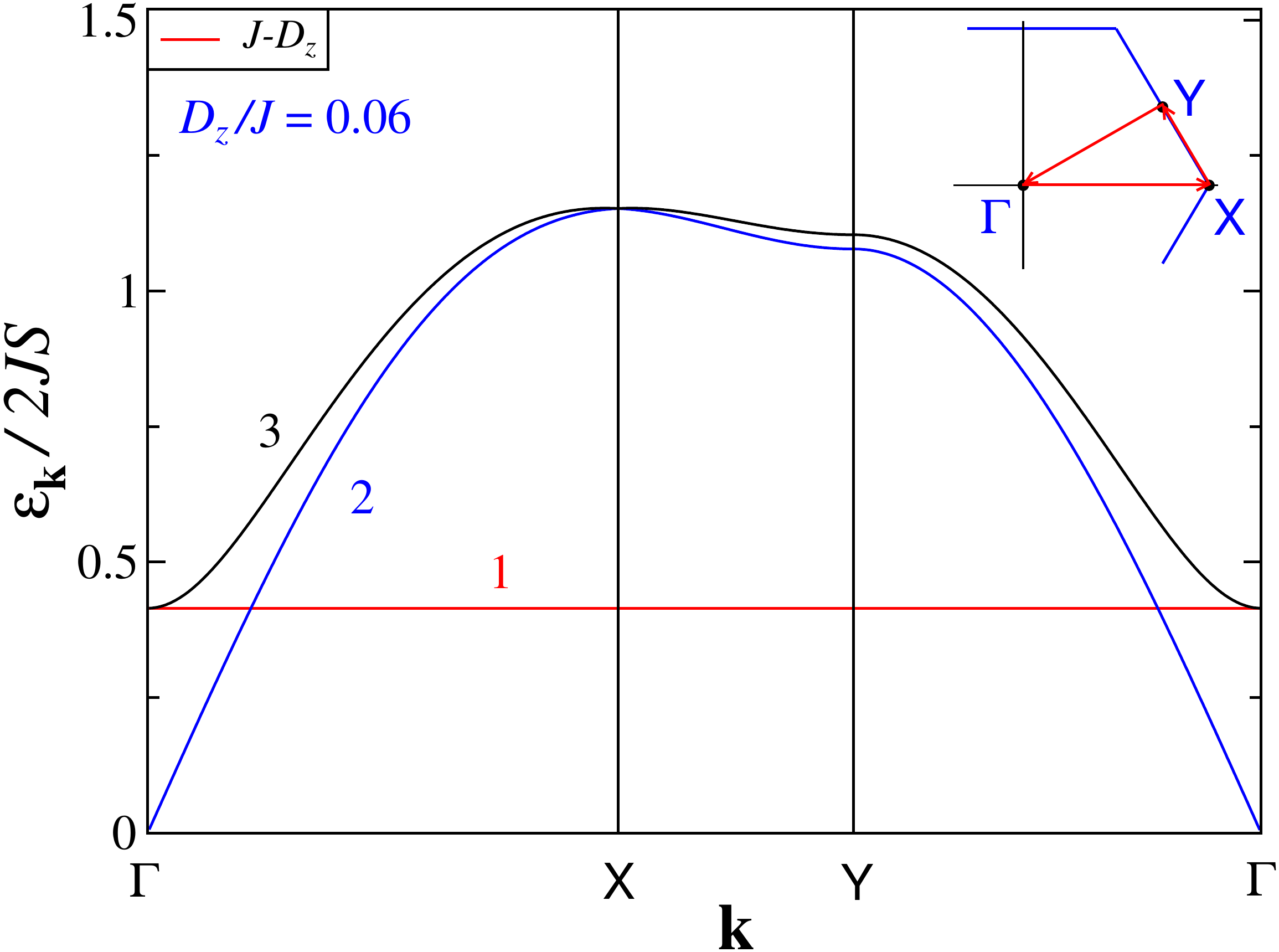}
\caption{(Color online)
Energies of the three magnon modes for the $J$--$D_z$ spin model with the out-of-plane DM 
interaction $D_z/J =0.06$ along a representative path in the Brillouin zone.
}
\label{Fig:wkDM}
\end{figure}

Since the DM term concerns only the nearest-neighbor pairs of spins and because 
in the ${\bf q}\!=\!0$  state all DM bonds contribute identically, the overall structure of the harmonic 
part of the Hamiltonian remains the same as in \eqref{H2F}.
Then, some algebra yields the harmonic Hamiltonian $\hat{\cal H}_{2}$ in the form 
\eqref{Hmatrix1} with  the DM contributions
\begin{eqnarray}
\delta\hat{\bf A}_{\bf k} = d_M \left( \hat{\bf I}-
\frac{ 1 }{4}\,\hat{\bm\Lambda}_{\bf k}\right),  \ \ \
\delta\hat{\bf B}_{\bf k} = \frac{ d_M }{4} \, \hat{\bm\Lambda}_{\bf k} \, ,
\label{ABDM}
\end{eqnarray}
where $d_M\!=\!\sqrt{3}D_z/J$ and $\hat{\bm\Lambda}_{\bf k}$ is unchanged from \eqref{Mk}, 
again reducing the eigenvalue problem of the harmonic spin-wave theory to the one 
of $\hat{\bm\Lambda}_{\bf k}$, already  solved in \eqref{lambda} and \eqref{lambda123}. 
Then, the spin-wave frequencies for the problem with the out-of-plane DM interaction 
\eqref{HDM} are $\omega_{\nu,\bf k}=\sqrt{\left(1+d_M\right)
\left(1-\lambda_{\nu,\bf k}/2\right)\left(1+d_M+\lambda_{\nu,\bf k}\right)}$.
With $\lambda_{\nu,\bf k}$ from \eqref{lambda123}, the flat mode energy  is
\begin{equation}
\label{w1DM}
\varepsilon_{1,\bf k}  = 2JS \sqrt{3d_M\left(1+d_M\right)/2} \, ,
\end{equation}
and the energies of the dispersive modes are
\begin{eqnarray}
\label{w23DM}
&&\varepsilon_{2(3),\bf k}= 2JS \sqrt{1+d_M}\\
&&\phantom{\varepsilon_{2(3),\bf k}}\times\sqrt{1+d_M-\gamma_{\bf k} -d_M
\bigl(1 \pm \sqrt{1 + 8\gamma_{\bf k}}\,\bigr)/4}\, ,\nonumber \ \ \ \ 
\end{eqnarray}
which are similar to the $XXZ$ \eqref{w1xxz} and the single-ion  
\eqref{w1si}, \eqref{w23si} results and are in agreement with Ref.~\onlinecite{Yildirim06}.
The energies of the magnon modes $\varepsilon_{\nu,\bf k}$ are illustrated 
in Fig.~\ref{Fig:wkDM} for the DM term  
$D_z/J\!=\!0.06$, a value relevant to the  Fe-jarosite.\cite{Matan06,Yildirim06}

Opposite to the previous extensions, the DM term contributes to the cubic anharmonicities, see \eqref{HDM12local}. 
Using the second term in \eqref{HDM12local} for one bond and extending it to 
the entire lattice we find for the ${\bf q}\!=\!0$ state
\begin{equation}
\hat{\cal H}_3 =\frac{d_M}{3}\, J\sqrt{\frac{S}{2}} \sum_{i,j} \sin\theta_{ij} \bigl( a_i^\dagger a_j^\dagger a^{\phantom{\dag}}_j +
\textrm{h.c.}\bigr)\, ,
\label{H3DM}
\end{equation}
whose structure is identical to the cubic term from the $J$-part of the Hamiltonian \eqref{H3s} 
considered in Sec.~\ref{Sec:nonlinear}. Thus, the out-of-plane DM interaction in the ${\bf q}\!=\!0$ state 
simply renormalizes cubic vertices  by a factor $(1+d_M/3)$.

\subsection{Small-$J_2$ expansion}

In the Heisenberg kagom\'e-lattice antiferromagnet, additional next-nearest-neighbor coupling $J_2$   
lifts the degeneracy of the 120$^\circ$ manifold of classical states and selects 
between the ${\bf q}\!=\!0$ and $\sqrt{3}\times\!\sqrt{3}$ ground states.\cite{Harris92,Chubukov92}  
It  also introduces a dispersion into the ``flat mode'' energy and thus was invoked to reproduce experimentally 
observed ${\bf k}$-dependence of such mode in Fe-jarosite.\cite{Yildirim06} 
We note, however, that quantum fluctuations beyond 
the harmonic order also generate effective $J_2$ interactions,\cite{Chubukov92,ChZh14}
and thus could be the source of the same ${\bf k}$-dependence. As we show in Sec.~\ref{Sec:nonlinear},
the dispersion of the flat mode  is particularly important for the quasi-resonant spin-wave decays. 
Below, we consider the effect of  small $J_2$ perturbatively.  Other types of small interactions
can be taken into account in a similar fashion.
 
We first point out  that the network of the second-neighbor bonds forms 
three independent kagom\'e lattices. Second, these bonds  connect only spins from 
different sublattices of the original lattice, see Fig.~\ref{basis}.
Therefore,   contribution of the $J_2$ term to the harmonic spin-wave Hamiltonian has the same structure 
as the nearest-neighbor Hamiltonian \eqref{H2F}  
\begin{eqnarray}
\delta\hat{\cal H}^\prime_2 &=& 2 J_2S \sum_{{\bf k},\alpha\beta}\biggl\{\Bigl[\delta_{\alpha \beta}+
\frac{1}{4}\,
\Lambda^{\prime \,\alpha\beta}_{\bf k} \Bigr] a_{\alpha,\bf k}^\dagger a^{\phantom{\dag}}_{\beta,\bf k} \nonumber\\
&&\phantom{2J_2S}- \frac{3}{8}\,\Lambda^{\prime\, \alpha\beta}_{\bf k} \bigl(
a_{\alpha,\bf k}^\dagger a^\dagger_{\beta,-\bf k}+\textrm{h.c.}\bigr)\biggr\},\label{H2J2}
\end{eqnarray}
where instead of $\hat{\bm\Lambda}_{\bf k}$ the matrix is 
\begin{equation}
\label{Lprime}
\hat{\bm\Lambda}^\prime_{\bf k} =\left(\begin{array}{ccc}
0   &  c^\prime_3 & c^\prime_1 \\
c^\prime_3 &  0   & c^\prime_2 \\
c^\prime_1 &  c^\prime_2 &  0
\end{array}\right) \, ,
\end{equation}
and we use the shorthand notations
$c^\prime_1 \!=\!\cos(q_3\!+\!q_2)$, $c^\prime_2 \!=\!\cos(q_3\!-\!q_1)$, 
$c^\prime_3\!=\!\cos(q_1\!+\!q_2)$
with $q_n\!=\!{\bf k}\cdot\bm{\delta}_n/2$.

Therefore, the diagonalization of the harmonic part of the $J-J_2$  Hamiltonian  
requires diagonalization of the matrix 
$\widetilde{\bm\Lambda}_{\bf k}\!=\!\hat{\bm\Lambda}_{\bf k}+j_2 \hat{\bm\Lambda}^\prime_{\bf k}$, 
where $j_2\!=\!J_2/J$. Generally, this can be done numerically, the approach taken  
in Ref.~\onlinecite{Yildirim06}  in the analysis of the Fe-jarosite spectrum, 
with the analytical results given only for the high-symmetry points. 
However, having in mind the  problem of spin-wave excitation in large-$S$ kagom\'e-lattice antiferromagnet,
for the physical range of interest $J_2\!\ll\!J$ one can make analytical progress using an expansion in $j_2$.

\begin{figure}[t]
\includegraphics[width=0.99\columnwidth]{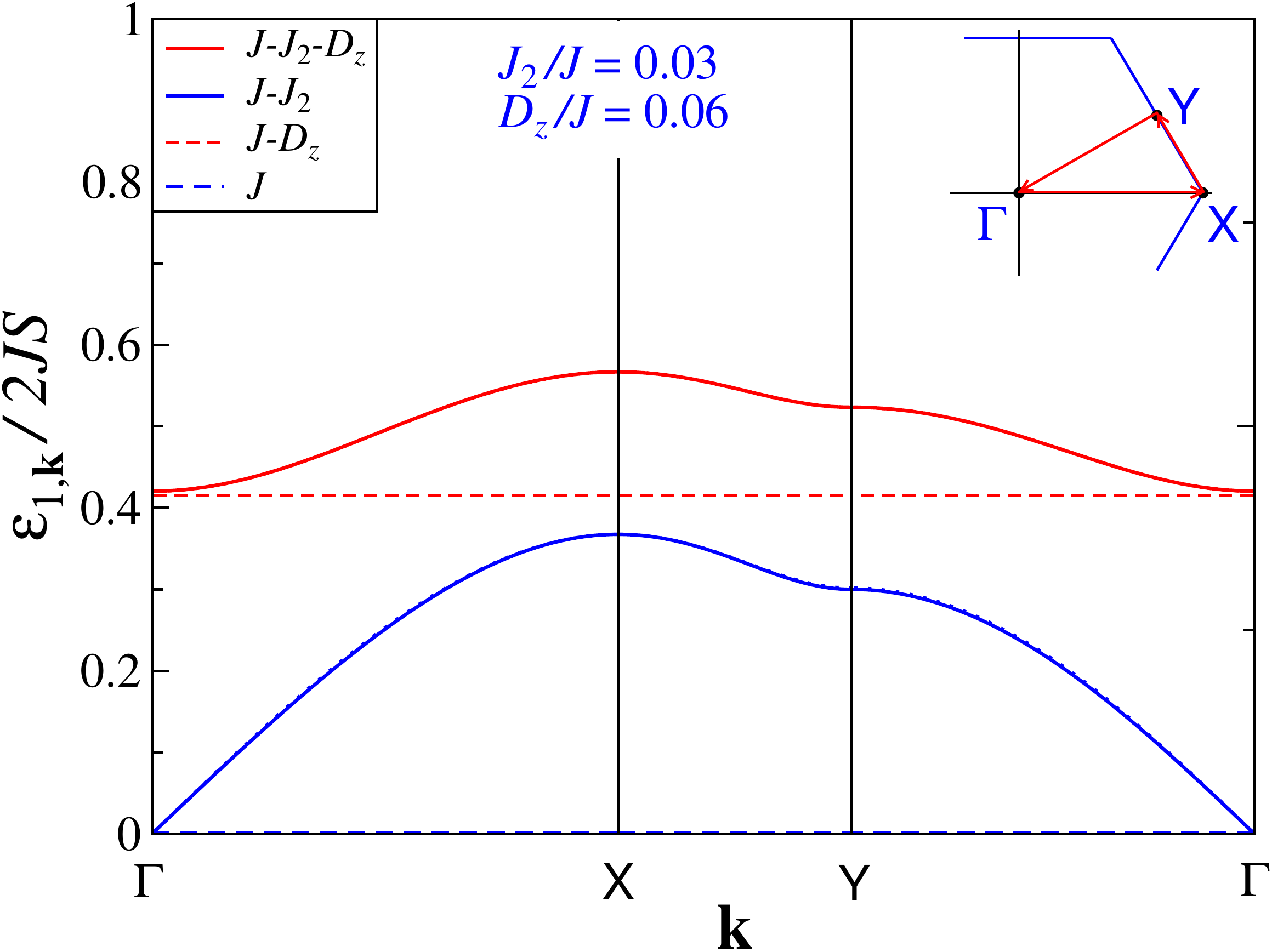}
\caption{(Color online)
Evolution of the ``flat mode'' upon switching on additional interactions with parameters shown in the legend.
Solid lines: energies of the ``flat mode''  for the $J-J_2$ Heisenberg (lower curve) and 
out-of-plane DM models (upper curve), using approximate expressions in Eqs.~(\ref{w1J1J2}) and (\ref{w1J1J2Dz}). 
Dashed lines are the energies for the same models at $J_2=0$. 
Dotted lines (virtually indistinguishable from solid lines) are exact results, without using the expansion in $J_2$.
}
\label{Fig:wk_J2}
\end{figure}

Because the most important qualitative effect of $J_2$ is the  dispersion of the flat band, 
we will ignore   small corrections to the ``normal'' modes.
Expanding characteristic equation for the matrix $\widetilde{\bm\Lambda}_{\bf k}$ in $j_2$ we obtain 
\begin{eqnarray}
&&\left(\lambda +1\right) \left(\lambda^2 - \lambda - 2\gamma_{\bf k}\right) = 
2j_2\left(\lambda f_1({\bf k})+f_2({\bf k})\right)   ,\ \ \
\label{lambdaprime} \\
&&\mbox{where}  \ \ \ 
f_1({\bf k})=c_1^\prime c_1+c_2^\prime c_2+c_3^\prime c_3\,,   \nonumber\\
&&\phantom{\mbox{where}  \ \ \ }
f_2({\bf k})=c_1^\prime c_2 c_3+c_2^\prime c_1 c_3 + c_3^\prime c_1 c_2  \, .\nonumber
\end{eqnarray}
Then,  the ``flat mode'' $\lambda_1\!=\!-1$ eigenvalue is modified as
\begin{equation}
\widetilde{\lambda}_{1,{\bf k}} = \lambda_1 +j_2 \lambda^{(1)}_{1,{\bf k}}\, , \ \ 
\lambda^{(1)}_{1,{\bf k}}=\left(\frac{f_2({\bf k})-f_1({\bf k})}{1-\gamma_{\bf k}}\right)  \, .
\label{lambda1prime}
\end{equation}
Corrections to $\lambda_2$ and $\lambda_3$ can be obtained  similarly.

Having this perturbative correction to  $\lambda_1$ in \eqref{lambda1prime} allows  us to obtain 
the flat mode energies in various models. We list some of the answers below.

\noindent
(i) $J-J_2$ Heisenberg model:
\begin{equation}
\label{w1J1J2}
\varepsilon_{1,\bf k}  = 2JS \sqrt{3j_2(1+\lambda^{(1)}_{1,{\bf k}})/2} + {\cal O}(j_2^{3/2})\, .
\end{equation}

\noindent
(ii) $J-J_2$ $XXZ$ model. Here we assume that the anisotropy $\Delta$ is the same in both exchanges: 
\begin{eqnarray}
\label{w1J1J2XXZ}
\varepsilon_{1,\bf k}  &=& 2JS \sqrt{3/2+j_2(1-\lambda^{(1)}_{1,{\bf k}}/2)}\\
&&\times\sqrt{1-\Delta+j_2(1+\lambda^{(1)}_{1,{\bf k}})} + {\cal O}(j_2^{2})\, .\nonumber
\end{eqnarray}

\noindent
(iii)  $J-J_2$ out-of-plane DM model.
\begin{eqnarray}
\label{w1J1J2Dz}
\varepsilon_{1,\bf k}  &=& 2JS \sqrt{3(1+d_M)/2+j_2
(1-\lambda^{(1)}_{1,{\bf k}}/2)} \ \ \ \\
&&\times\sqrt{d_M+j_2(1+\lambda^{(1)}_{1,{\bf k}})} + {\cal O}(j_2^{2})\, ,\nonumber
\end{eqnarray}
where $d_M\!=\!\sqrt{3}D_z/J$ as before. This is the model which was used to describe the spectrum of  
Fe-jarosite.\cite{Yildirim06}
Our results agree exactly with the expressions for the high-symmetry points provided in Ref.~\onlinecite{Yildirim06}.  
The advantage of our approach is that it is  fully analytical in the entire Brillouin zone. 

Effects of the second-neighbor exchange $J_2$ and the DM coupling $D_z$ on the 
dispersion of the ``flat mode'' $\varepsilon_{1,\bf k}$ 
are summarized in Fig.~\ref{Fig:wk_J2} for representative values $J_2/J\!=\!0.03$ and $D_z/J\!=\!0.06$
that are motivated by the experimental data for Fe-jarosite; \cite{Matan06,Yildirim06}
the $J_2\!=\!0$ energies are also shown. In the same figure the energies obtained via 
numerical diagonalization of  the matrix $\widetilde{\bm\Lambda}_{\bf k}$ are shown. 
The corresponding results are indistinguishable from the approximate results of Eqs.~(\ref{w1J1J2}) and (\ref{w1J1J2Dz})
 on the scale of our plot.

\subsection{Two-step diagonalization}

For the above three cases, 
the $XXZ$, the single-ion, and the
out-of-plane DM models, the harmonic spin-wave theory includes
diagonalization of the same matrix $\hat{\bm\Lambda}_{\bf k}$. Here we elaborate on details of 
this general procedure and provide the formalism, which is essential for treating the non-linear terms in all these models  
and is identical for all  considered cases.

Following Ref.~\onlinecite{Harris92},  diagonalization of
$\hat{\bm\Lambda}_{\bf k}$ implies a two-step diagonalization procedure
of $\hat{\cal H}_2$ in the form (\ref{Hmatrix1}).
Its eigenvectors ${\bf w}^\dagger_{\nu}=\left(w_{\nu,1}({\bf k}),w_{\nu,2}({\bf k}),w_{\nu,3}({\bf k})\right)$ obey
\begin{equation}
\hat{\bm\Lambda}_{\bf k} {\bf w}_{\nu} =\lambda_{\nu,\bf k} {\bf w}_{\nu}
\end{equation}
and can be found explicitly\cite{Harris92,ChZh14} 
\begin{equation}
\label{wn}
{\bf w}_\nu ({\bf k})= \frac{1}{r_\nu} \,\left(\begin{array}{c}
c_1c_2 + \lambda_\nu c_3 \\
\lambda_\nu^2 - c_1^2 \\
c_1 c_3 + \lambda_\nu c_2
\end{array}\right)\, ,
\end{equation}
with $r_\nu = \sqrt{(c_1c_2 + \lambda_\nu c_3)^2\! +
(\lambda_\nu^2 - c_1^2)^2 \! + (c_1 c_3\! + \lambda_\nu c_2)^2}$.

These eigenvectors define a unitary transformation of the original Holstein-Primakoff
bosons  to the new ones
\begin{equation}
a^{\phantom{\dag}}_{\alpha,\bf k} = \sum_\nu w_{\nu,\alpha}({\bf k})\, d^{\phantom{\dag}}_{\nu,\bf k}\,  , \ \
d^{\phantom{\dag}}_{\nu,\bf k} = \sum_\alpha w_{\nu,\alpha}({\bf k})\, a^{\phantom{\dag}}_{\alpha,\bf k}\, ,
\label{linearT}
\end{equation}
such that  the harmonic Hamiltonian $\hat{\cal H}_2$ in the form of (\ref{H2F})  with 
$\hat{\bf A}_{\bf k}$'s and $\hat{\bf B}_{\bf k}$'s from (\ref{AB}) with (\ref{ABxxz}), (\ref{ABsi}), or (\ref{ABDM})  
is turned into three independent Hamiltonians 
$$
\hat{\cal H}_2 = 2JS  \sum_{\nu,\bf k} \Bigl[ A_{\nu,\bf k} d_{\nu,\bf k}^\dagger d^{\phantom{\dag}}_{\nu,\bf k} 
 - \frac{B_{\nu,\bf k}}{2}\bigl(d^\dagger_{\nu,\bf k} d^\dagger_{\nu,-\bf k} + \textrm{h.\,c.}\bigr)\Bigr].
$$
The final step   is the conventional   
Bogolyubov transformation for each individual species of the $d$-bosons
\begin{equation}
d^{\phantom{\dag}}_{\nu,\bf k} = u_{\nu\bf k}  b^{\phantom{\dag}}_{\nu,\bf k} + v_{\nu\bf k}  b^\dagger_{\nu,-\bf k} \,,
\label{BogolyubovT}
\end{equation}
with $u_{\nu\bf k}^2 -  v_{\nu\bf k}^2 = 1$ and 
\begin{equation}
v_{\nu\bf k}^2= \frac12\Bigl(\frac{A_{\nu,\bf k}}{\omega_{\nu,\bf k}}-1\Bigr),\quad
2u_{\nu\bf k}v_{\nu\bf k} = \frac{B_{\nu,\bf k}}{\omega_{\nu,\bf k}} \,,
\label{Bogolyubov_uv}
\end{equation}
where the eigenvalues $\omega_{{\nu,\bf k}}$, $A_{\nu,\bf k}$, and $B_{\nu,\bf k}$ were obtained for each of the models in 
previous sections.
The importance of this two-step procedure will be apparent in the discussions of the non-linear terms in Sec.~\ref{Sec:nonlinear}.

\subsection{Ordered magnetic moment}

Here we discuss the dependence of the ordered magnetic moment on anisotropy parameter and on the value of the 
spin $S$. While we only consider the $XXZ$ model,\cite{ChZh14} the results are expected to be 
similar for the other anisotropies considered above.
It should be noted that this calculation can only qualitatively  estimate the  stability of the Ne\'{e}l order, because
it only includes the ``diagonal'' quantum fluctuation for a given state, completely neglecting  
the ``off-diagonal''  tunneling  processes between different states within the
manifold. On the other hand, such processes should be exponentially suppressed for larger spins. \cite{Henley93}

Within the linear spin-wave theory, magnetic moment on a site that belongs
to the sublattice $\alpha$ is reduced by zero-point fluctuations:
$\langle S\rangle_\alpha=S-\langle a_{\alpha,\ell}^\dag a^{\phantom{\dag}}_{\alpha,\ell}\rangle$.
Converting  $a_\alpha$ to $d_\mu$ and then to $b_\mu$  using unitary (\ref{linearT}) and
 Bogolyubov (\ref{BogolyubovT}) transformations  one arrives to
\begin{equation}
\langle S\rangle_\alpha=S-\frac{1}{N}\sum_{\mu,{\bf k}}w^2_{\mu,\alpha}({\bf k}) \, v^2_{\mu{\bf k}}\,.
\label{Sav1}
\end{equation}
Since all three sublattices are equivalent, symmetrization of (\ref{Sav1}) gives
\begin{equation}
\langle S\rangle=S-\frac{1}{3N}\sum_{\mu,{\bf k}} v^2_{\mu{\bf k}}\,,
\label{Sav2}
\end{equation}
with $v^2_{\mu{\bf k}}$ from (\ref{Bogolyubov_uv}). 
Calculations of the magnetization $M\!=\!\langle S\rangle/S$ and the $\langle S\rangle\!=\!0$
Ne\'{e}l order boundary in the $S\!-\!\Delta$ plane are performed taking the 2D integrals in (\ref{Sav2})  numerically
 using (\ref{Bogolyubov_uv}) for the $XXZ$ model.

\begin{figure}[t]
\includegraphics[width=0.8\columnwidth]{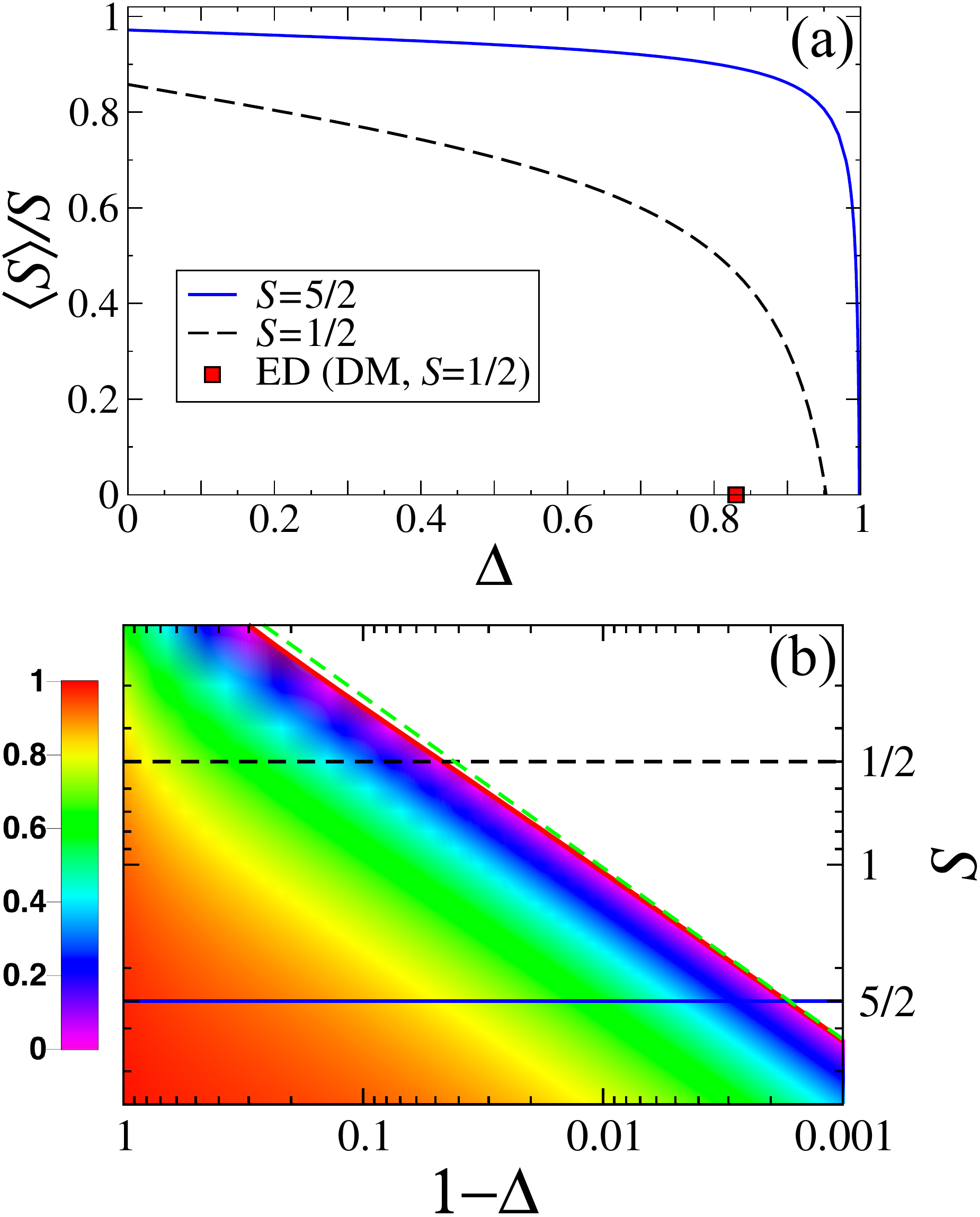}
\caption{(Color online)\ (a)  Magnetization 
$M\!=\!\langle S\rangle/S$ vs $\Delta$  for $S\!=\!1/2$ and $S\!=\!5/2$.
Square is the ED result for $S\!=\!1/2$ with the DM interaction.
(b) Intensity plot of the magnitude of $M$. Solid line is the $\langle S\rangle\!=\!0$ Ne\'{e}l order phase 
boundary in the $S\!-\!\Delta$ plane on the log-log scale, dashed line
is the asymptotic approximation for it, see text.
}
\label{deltaS}
\end{figure}

Quantum suppression of the ordered moment  vs anisotropy $\Delta$ is shown in Fig.~\ref{deltaS}(a)   
for two values of spin. Linear spin-wave theory suggests a disordered state near the Heisenberg limit for all spin values
because quantum correction diverges for $\Delta \to 1$ due to vanishing energy
of the ``flat mode.''
The critical value $1\!-\!\Delta_c \!\approx\! 0.047$ for $S\!=\!1/2$ is compared  in Fig.~\ref{deltaS}(a)  with the result for
the DM coupling $D_{z,c} \!=\!0.1$ (rescaling $1\!-\!\Delta_c\!\Leftrightarrow\!\sqrt{3}D_{z,c}$),
found by exact diagonalization (ED).  \cite{Cepas08,Messio10}
Since the DM term suppresses tunneling processes within the manifold,  it is reasonable
to compare ED with spin-wave theory results to evaluate the accuracy of the  Ne\'{e}l order boundary.
One can see a qualitative agreement with ED and a quantitative exaggeration of the
extent of the ordered phase, expected for the spin-wave approach.

Near the Heisenberg limit, one can neglect non-divergent terms in   (\ref{Sav2}) and find an
asymptotic expression for the Ne\'{e}l order boundary from $\langle S\rangle\!=\!0\!\approx\! S\!-\!A_1/6\omega_1$,
where $A_1=3/4$, see (\ref{AB1}), and $\omega_1=\sqrt{3(1-\Delta)/2}$, see (\ref{w1xxz}), leading to
$1\!-\!\Delta_c\!\approx\! 1/96S^2$,
which is shown in Fig.~\ref{deltaS}(b) together with the result of the numerical integration in (\ref{Sav2})
and demonstrates an exceedingly close agreement with it.

\section{Non-linear Spin-wave theory}
\label{Sec:nonlinear}

In Sec.~\ref{Sec:linear} we have considered three anisotropic models of the 
kagom\'e-lattice antiferromagnets in the harmonic  approximation and  outlined the 
approach to taking into account other terms, such as further-neighbor exchanges, 
perturbatively. In this Section, we derive the nonlinear, cubic and quartic terms 
of the spin-wave $1/S$ expansion  and then exemplify their role in the ground-state 
selection and in the spectral properties of the kagom\'e-lattice antiferromagnets 
using representative cases. 

Below, we first obtain cubic and quartic terms of the spin-wave expansion to conclude the 
formal development of the theory. Then the cubic vertices for   
${\bf q}\!=\!0$ and $\sqrt{3}\times\!\sqrt{3}$ states  allow  us to  
proceed with calculating order-by-disorder fluctuating corrections to their ground-state 
energies for the $XXZ$ and single-ion anisotropy models. Both models 
demonstrate a quantum phase transition between the ${\bf q}\!=\!0$ and the $\sqrt{3}\times\!\sqrt{3}$ states
as a function of  anisotropy parameter. Hence, both cases present    rare examples of the 
quantum order-by-disorder favoring a different state  from the one selected by  thermal fluctuations,
the latter choosing the $\sqrt{3}\times\!\sqrt{3}$  structure regardless of the anisotropy. 
\cite{Harris92,Reimers93,Henley09,Huse92,Korshunov02,Rzchowski97,ChZh14,Gotze15} 

We then proceed with a calculation of the decay-induced effects in the structure-factor ${\cal S}({\bf q},\omega)$
within the  DM model with $J_2$. While this calculation is aimed at  giving a detailed account of the 
spectral properties of a specific kagom\'{e}-lattice antiferromagnet described by this model, $S=5/2$ Fe-jarosite, 
the outlined scenario should be applicable to a wide variety of the other flat-band frustrated spin systems.
\cite{Petrenko,Matan14,Balents_honeycomb,Georgii15} 
We also note that the structure factor ${\bf q}$-dependence allows to ``filter out'' spectral
contributions of specific  modes  in the portions of 
the ${\bf q}$-space  while highlighting the others; a useful phenomenon characteristic 
to the non-Bravais lattices.

\subsection{Cubic terms}

In the $XXZ$, single-ion anisotropy, and Heisenberg  models, (\ref{Hs}) with or without the anisotropies in 
(\ref{Hsxxz}) and (\ref{HD}), the terms that lead to the 
cubic anharmonic coupling of the spin waves are identical and originate from the 
$S_i^x S_j^z$ part of  (\ref{Hs}). These terms are also the only ones
that are able to distinguish between different 120$^\circ$ spin configurations in these models
by virtue of containing  $\sin\theta_{ij}\!= \!\pm\sqrt{3}/2$ for the clockwise
or counterclockwise spin rotation.
In the bosonic representation they yield  
\begin{equation}
\hat{\cal H}_3 = J\sqrt{\frac{S}{2}} \sum_{i,j} \sin\theta_{ij} \bigl( a_i^\dagger a_j^\dagger a_j +
\textrm{h.c.}\bigr)\, ,
\label{H3s}
\end{equation}
where $\theta_{ij}= \theta_i - \theta_j$ is the angle between two neighboring spins as before.
Clearly, the spin-wave interaction resulting from this term has different amplitude for different spin patterns.
For the DM model, the DM term  (\ref{HDM}) also contributes to the cubic anharmonicity, but its 
structure for the ${\bf q}\!=\!0$ state is identical to (\ref{H3s}), see  (\ref{H3DM}), so it gives a 
simple change of the overall factor in the vertex 
$J\!\rightarrow\!J\!+\!D_z/\sqrt{3}$.  

Using (\ref{H3s}), we now detail the derivation of the  cubic vertices for the main contenders for the ground state, the
${\bf q}\!=\!0$ and the $\sqrt{3}\times\!\sqrt{3}$ states.
We begin with the ${\bf q}\!=\!0$  pattern, for which $\hat{\cal H}_3$ in (\ref{H3s}) can be rewritten as
\begin{equation}
\hat{\cal H}_3= -J\sqrt{\frac{3S}{2N}}\sum_{\alpha\beta,\bf k,q}
\epsilon^{\alpha\beta\gamma} \cos(q_{\beta\alpha})
a^\dagger_{\alpha,\bf q} a^\dagger_{\beta,\bf k} a_{\beta,\bf p}+ \textrm{h.\,c.},
\label{H30}
\end{equation}
where $\epsilon^{\alpha\beta\gamma}$ is the Levi-Civita
antisymmetric tensor, ${\bf p}={\bf k}+{\bf q}$, 
$q_{\beta\alpha}={\bf q}\cdot\bm{\rho}_{\beta\alpha}$,
 and $\bm{\rho}_{\beta\alpha} = \bm{\rho}_\beta - \bm{\rho}_\alpha$.

The unitary transformation (\ref{linearT}) in (\ref{H30})  yields
\begin{equation}
\hat{\cal H}_3 = -J\sqrt{\frac{3S}{2N}}\sum_{\bf k,q}\sum_{\nu\mu\eta}
F^{\nu\mu\eta}_{\bf qkp}\,
d_{\nu,\bf q}^\dagger d_{\mu,\bf k}^\dagger  d_{\eta,\bf p}  + \textrm{h.c.},
\label{H31}
\end{equation}
where ${\bf p}={\bf k}+{\bf q}$ and the amplitude is 
\begin{equation}
F^{\nu\mu\eta}_{\bf qkp}=  \sum_{\alpha\beta}
\epsilon^{\alpha\beta\gamma}\cos(q_{\beta\alpha}) \,
w_{\nu,\alpha}({\bf q}) w_{\mu,\beta}({\bf k}) w_{\eta,\beta}({\bf p}).
\label{F0}
\end{equation}
Finally, the Bogolyubov transformation (\ref{BogolyubovT}) gives
\begin{eqnarray}
&&\hat{\cal H}_3  = \frac{1}{3!} \frac{1}{\sqrt{N}} \sum_{\bf k,q}\sum_{\nu\mu\eta}
V^{\nu\mu\eta}_{\bf qkp}\,
b_{\nu,\bf q}^\dagger b_{\mu,\bf k}^\dagger  b^\dagger_{\eta,\bf -p} + \textrm{h.c.}, \quad\quad
\label{Hss}\\
&&\phantom{\hat{\cal H}_3\ \ } + \frac{1}{2!} \frac{1}{\sqrt{N}} \sum_{\bf k,q}\sum_{\nu\mu\eta}
\Phi^{\nu\mu\eta}_{\bf qk;p}\,
b_{\nu,\bf q}^\dagger b_{\mu,\bf k}^\dagger  b_{\eta,\bf p} + \textrm{h.c.}, \ \ \ \ \ 
\label{Hd}
\end{eqnarray}
with the  vertices for  the ``source'' and the ``decay'' terms  
\begin{eqnarray}
V^{\nu\mu\eta}_{\bf qkp}= -J\sqrt{\frac{3S}{2}}\;
\widetilde{V}^{\nu\mu\eta}_{\bf qkp} \,,\ \ 
\Phi^{\nu\mu\eta}_{\bf qk;p}= -J\sqrt{\frac{3S}{2}}\;
\widetilde{\Phi}^{\nu\mu\eta}_{\bf qk;p} \,,
\label{V30}
\end{eqnarray}
where the symmetrized dimensionless vertices are  
\begin{eqnarray}
\widetilde{V}^{\nu\mu\eta}_{\bf qkp} & = & 
F^{\nu\mu\eta}_{\bf qkp} (u_{\nu\bf q}+v_{\nu\bf q})(u_{\mu\bf k}v_{\eta\bf p}+v_{\mu\bf k}u_{\eta\bf p})
\nonumber\\
&& \mbox{} +
F^{\mu\eta\nu}_{\bf kpq} (u_{\mu\bf k}+v_{\mu\bf k})(u_{\nu\bf p}v_{\eta\bf q}+v_{\nu\bf p}u_{\eta\bf q})
\qquad
\label{V3}\\
&& \mbox{} +
F^{\eta\nu\mu}_{\bf pqk} (u_{\eta\bf p}+v_{\eta\bf p})(u_{\nu\bf q}v_{\mu\bf k}+v_{\nu\bf q}u_{\mu\bf k})\,,
\nonumber
\end{eqnarray}
and
\begin{eqnarray}
\widetilde{\Phi}^{\nu\mu\eta}_{\bf qk;p}  & = & 
F^{\nu\mu\eta}_{\bf qkp} (u_{\nu\bf q}+v_{\nu\bf q})(u_{\mu\bf k}u_{\eta\bf p}+v_{\mu\bf k}v_{\eta\bf p})
\nonumber\\
&& \mbox{} +
F^{\mu\eta\nu}_{\bf kpq} (u_{\mu\bf k}+v_{\mu\bf k})(u_{\nu\bf p}u_{\eta\bf q}+v_{\nu\bf p}v_{\eta\bf q})
\qquad
\label{V3d1}\\
&& \mbox{}  +
F^{\eta\nu\mu}_{\bf pqk} (u_{\eta\bf p}+v_{\eta\bf p})(u_{\nu\bf q}v_{\mu\bf k}+v_{\nu\bf q}u_{\mu\bf k})\,,
\nonumber
\end{eqnarray}
where we have used the symmetry property $F^{\nu\mu\eta}_{\bf qkp}\!=\!F^{\nu\eta\mu}_{\bf qpk}$.

Repeating the same calculation for the $\sqrt{3}\times\sqrt{3}$ state we obtain identical expressions
for the cubic spin-wave Hamiltonian  and corresponding vertices, but
with different amplitudes $F^{\nu\mu\eta}_{\bf qkp}$ expressed as
\begin{equation}
F^{\nu\mu\eta}_{\bf qkp} = i\sum_{\alpha\beta}
\epsilon^{\alpha\beta\gamma}\sin(q_{\beta\alpha}) \,
w_{\nu,\alpha}({\bf q}) w_{\mu,\beta}({\bf k}) w_{\eta,\beta}({\bf p}).
\label{F3}
\end{equation}
Explicit expressions for the unitary 
transformation eigenvectors $w_{\nu,\alpha}({\bf q})$ 
of the matrix $\hat{\bm\Lambda}_{\bf k}$ and of the parameters of the Bogolyubov transformation 
are instrumental in deriving analytic expressions of the cubic anharmonic terms.
We also note that for all  models considered here, the functional form of the cubic
vertices (\ref{V3}) and (\ref{V3d1}) is identical, with all the differences  hidden  in the expressions 
of the Bogolyubov parameters $u_{\mu{\bf k}}$ and $v_{\mu{\bf k}}$ from (\ref{Bogolyubov_uv}). 

The role of the cubic terms in the ground-state selection and in the spectrum of the kagom\'{e}-lattice
antiferromagnets is discussed below in Sec.~\ref{Sec:gs} and Sec.~\ref{Sec:spectrum}.

\subsection{Quartic terms}

In the Holstein-Primakoff bosonic representation of   spin models, quartic terms  
originate from the  $S_i^x S_j^x$, $S_i^y S_j^y$, and $S_i^z S_j^z$  parts of the Hamiltonian.
In the models considered here, quartic terms do not help to differentiate between different states of the 120$^\circ$
manifold, but lead to the overall ground-state energy shift and   to the Hartree-Fock corrections to the 
spin-wave energies. Because of the coplanar 120$^\circ$ spin configuration, the formal expressions for 
these contribution show a close similarity to the ones for the triangular-lattice Heisenberg 
model, see Ref.~\onlinecite{triPRB09}.
Therefore,   we simply list the quartic parts of the  Hamiltonians of the $XXZ$ model and the DM term
together with the expressions for the ground-state energy shift in the former model and for the spin-wave energy 
correction for the latter model. Technical details are offered in Appendix~\ref{AppA}.

The quartic terms in the $XXZ$ model, (\ref{Hs}) and (\ref{Hsxxz}), are
\begin{eqnarray}
&&\hat{\cal H}_{4}  =\frac{J}{2} \sum_{\langle ij\rangle}\: 
\biggl( \frac{2\Delta +1}{8} \bigl((n_i+n_j) a_i a_j 
+  {\rm h.c.} \bigr)  
\label{Hxxz4} \\
 & & \phantom{\hat{\cal H}_{4}  =\frac{J}{2}} 
 -\frac{2\Delta -1}{8}\:\bigl(a^\dagger_j (n_i+n_j)  a_i +  {\rm h.c.} \bigr)- n_i n_j \biggr),\nonumber
\end{eqnarray}
where $n_i=a^\dagger_i a_i$ and we omitted the $\alpha$ indices of the bosonic variables for brevity.

The quartic contribution from the DM term (\ref{HDM}) is
\begin{eqnarray}
&&\hat{\cal H}_{4,DM}  =\frac{D_z\sqrt{3}}{2} \sum_{\langle ij\rangle}\: 
\biggl( \frac{1}{8} \bigl((n_i+n_j) a_i a_j 
+  {\rm h.c.} \bigr)  
\label{HDM4}
 \\
 & & \phantom{\hat{\cal H}_{4,DM}  =\frac{D_z\sqrt{3}}{2}} 
 +\frac{1}{8}\:\bigl(a^\dagger_j (n_i+n_j)  a_i +  {\rm h.c.} \bigr)- n_i n_j \biggr).\nonumber
\end{eqnarray}
By means of the  Hartree-Fock decoupling\cite{Oguchi60} outlined in Appendix~\ref{AppA},
one can obtain  contribution of the quartic terms to the  
$1/S$ series expansion of the ground-state energy of any non-collinear spin model
\begin{eqnarray}
E=E_{\rm cl}+\delta E^{(2)}+\delta E^{(3)}+\delta E^{(4)},
\label{E}
\end{eqnarray}
where the first term is the classical energy of order ${\cal O}(S^2)$, the second is the
harmonic correction from $\hat{\cal H}_{2}$, ${\cal O}(S)$, and the last two are from the nonlinear cubic 
and  quartic terms, both ${\cal O}(1)$. While we discuss the cubic part of this expression later, 
the classical and harmonic energy terms (per spin) of the series in (\ref{E}) for the $XXZ$ model are
\begin{eqnarray}
E_{\rm cl}+\delta E^{(2)}=-JS^2-JS\bigg(1-\frac{1}{3N}\sum_{\mu,{\bf k}} \omega_{\mu,{\bf k}}\bigg),
\label{E2}
\end{eqnarray}
with the frequencies from (\ref{w1xxz}). The quartic member of the series (\ref{E}) for the same model is 
\begin{eqnarray}
&&\delta E_{4} = -J\Big[ n^2 +  m^2 + \bar\Delta^2 
\label{E4}\\
&&\
- (2\Delta+1)\Big(n\bar{\Delta} +  \frac{m\delta}{2}\Big)  
+ (2\Delta-1)\Big(nm + \frac{\bar{\Delta}\delta}{2} \Big)\Big],\nonumber
\end{eqnarray}
where the Hartree-Fock averages are introduced in a standard manner
\begin{equation}
n = \langle a^\dagger_i a_i\rangle , \ \ m = \langle a^\dagger_i a_j\rangle, 
\ \ \bar\Delta = \langle a_i a_j\rangle , \ \ 
\delta = \langle a^2_i\rangle,
\label{HF}
\end{equation}
and are given in Appendix \ref{AppA}.

In the same order of expansion, quartic terms lead to the Hartree-Fock 
corrections to the linear spin-wave Hamiltonian 
$$
\delta\hat{\cal H}_2  = 2J  \sum_{\nu,\bf k} \delta A_{\nu,\bf k} d_{\nu,\bf k}^\dagger d^{\phantom{\dag}}_{\nu,\bf k} 
 - \frac{\delta B_{\nu,\bf k}}{2}\bigl(d^\dagger_{\nu,\bf k} d^\dagger_{\nu,-\bf k} + \textrm{h.\,c.}\bigr),
$$
which yields the  Hartree-Fock part of the $1/S$ contribution to the spin-wave energy 
\begin{eqnarray}
\varepsilon^{(4)}_{\nu,\bf k} =2J\: 
\frac{ A_{\nu,\bf k}\delta A_{\nu,\bf k}- B_{\nu,\bf k}\delta B_{\nu,\bf k}}{\omega_{\nu,\bf k}}\,. 
\label{EHF4}
\end{eqnarray}
Here we give expressions for $\delta A_{\nu,\bf k}$ and $\delta B_{\nu,\bf k}$ for the Heisenberg model (\ref{Hs}) with the 
DM anisotropy (\ref{HDM}) as an example
\begin{eqnarray}
\delta A_{\nu,\bf k} =   \widetilde{C}_1+\frac{\widetilde{C}_2\lambda_{\nu,\bf k}}{2} ,
\ \
\delta B_{\nu,\bf k} = -  2\widetilde{C}_4-\frac{\widetilde{C}_3\lambda_{\nu,\bf k}}{2} , \label{H24a}
\end{eqnarray}
where $\widetilde{C}_i=C_i-d_M D_i$ and the  constants are the linear combinations of the binary Hartree-Fock
averages in (\ref{HF})  and can be found in Appendix \ref{AppA}.
We note that since the ``flat-mode'' eigenvalue $\lambda_{1,\bf k}=-1$, the quartic $1/S$ correction to its 
energy, $\varepsilon^{(4)}_{1,\bf k}$ in (\ref{EHF4}), is also necessarily momentum-independent 
in all   models considered here. Therefore, it is
a contribution of the cubic terms  which is going to introduce a fluctuation-induced dispersion
in the flat mode in the same order of the $1/S$ expansion.

\subsection{Ground state selection}
\label{Sec:gs}

Here, we discuss the role of cubic terms (\ref{Hss}) in the ground-state selection in the $XXZ$ and 
the single-ion anisotropy models. As was mentioned above, in the case of these  models, 
(\ref{Hs}) with (\ref{Hsxxz}) and (\ref{HD}), classical and harmonic terms are not able to lift the degeneracy 
in the manifold of  coplanar 120$^\circ$ structures. It is also easy to see from (\ref{Hs}) that the quartic terms 
are also unable to differentiate between 120$^\circ$ states, leaving the cubic term as a sole 
source of the quantum order-by-disorder effect to this order in $1/S$.
The same is also true for the single-ion term in (\ref{HD}).

\begin{figure}[t]
\centerline{
\includegraphics[width=0.99\columnwidth]{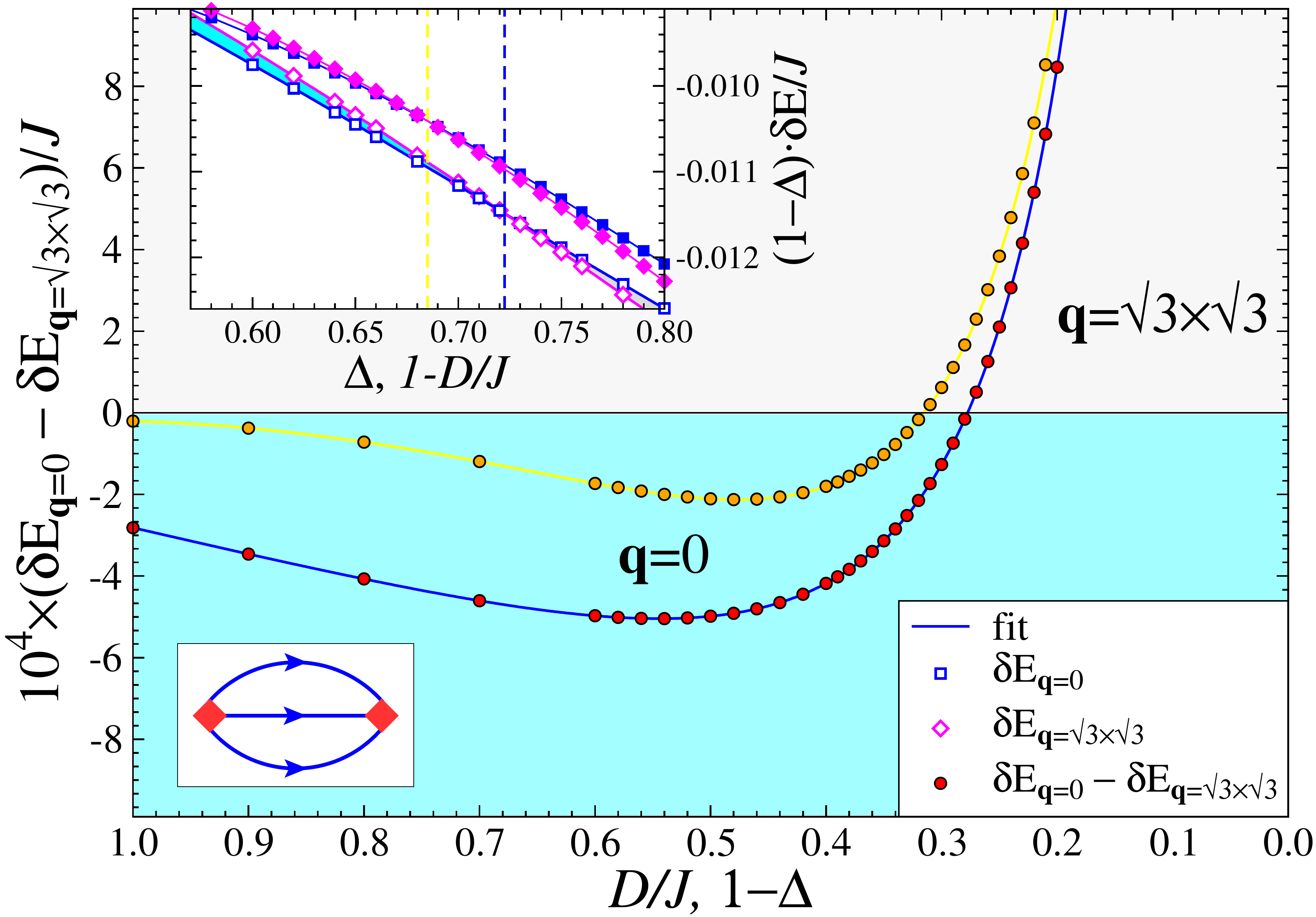}
}
\caption{(Color online) Difference of the ground-state energies 
of the ${\bf q}\!=\!0$ and $\sqrt{3}\times\!\sqrt{3}$ states  per spin. 
Upper inset:  energy correction $\delta E^{(3)}$
for the ${\bf q}\!=\!0$ (squares) and $\sqrt{3}\times\!\sqrt{3}$ state (diamonds). Dashed vertical lines mark
the transition. Upper data points/lines in the  figure and inset are 
for the single-ion anisotropy model, (\ref{Hs}) with (\ref{HD}), and 
the lower are for the $XXZ$ model, (\ref{Hs}) with (\ref{Hsxxz}).
Lower inset: diagram for  $\delta E^{(3)}$ term  (\ref{dE3s}) in the energy expansion (\ref{E}). 
}
\label{fig:dE3}
\end{figure}

The second-order energy correction from the cubic terms (\ref{Hss}) resulting in  $\delta E^{(3)}$ in (\ref{E})
is represented by the diagram in the lower inset of Fig.~\ref{fig:dE3} and is given by
\begin{eqnarray}
\delta E^{(3)}=-\frac{1}{18N^2}\sum_{\nu\mu\eta}\sum_{{\bf q},{\bf k}}
\frac{|V^{\nu\mu\eta}_{{\bf q},{\bf k},-{\bf k}-{\bf q}}|^2}
{\varepsilon_{\nu,\bf q} + \varepsilon_{\mu,\bf k} + \varepsilon_{\eta,-{\bf k}-{\bf q}}}\, ,
\label{dE3s}
\end{eqnarray}
where the energy is per spin and $N$ is the number of unit cells.
Summation over magnon branches $\mu,\nu,\eta$ gives twenty seven individual contributions
of which  ten are distinct. With the formal expression for the source vertices in (\ref{V3}), the energy correction
in (\ref{dE3s}) is identical for the $XXZ$ and the single-ion cases, with 
the  differences in the expressions for the spin-wave energies $\varepsilon_{\alpha,\bf k}$ in (\ref{w1xxz}) 
and (\ref{w1si}) and (\ref{w23si}), 
and in the $u_{\alpha\bf k}$ and $v_{\alpha\bf k}$ parameters (\ref{Bogolyubov_uv})   in vertices (\ref{V3}) for the corresponding
models.

Using an explicit form for the cubic vertices for the ${\bf q}\!=\!0$ and $\sqrt{3}\times\!\sqrt{3}$ 
spin states, we performed a high-accuracy numerical integration 
in Eq.~(\ref{dE3s}) and studied the quantum order-by-disorder effect.
Results of these calculations are presented in Fig.~\ref{fig:dE3}.
The quantum selection of the ${\bf q}\!=\!0$ state over the $\sqrt{3}\times\!\sqrt{3}$
counterpart for the large planar anisotropy $(1-\Delta_c) [D_c/J]\!\agt\!0.3$ was highlighted in our previous work
on the $XXZ$ model. \cite{ChZh14}
This qualitative conclusion is in contrast with the usual behavior, in which quantum fluctuations  lead to the 
same ground state that is selected by the  thermal fluctuations.
Indeed, for the classical kagom\'e-lattice antiferromagnets in both   Heisenberg and   $XY$  limits,
thermal fluctuations select the $\sqrt{3}\times\!\sqrt{3}$ magnetic structure
as the leading instability \cite{Harris92,Reimers93,Henley09,Huse92,Korshunov02,Rzchowski97} 
contrary to the behavior of the quantum model.

Here we complement the above result by the analysis of the single-ion anisotropy model (\ref{HD}). While 
the overall trend in $\delta E^{(3)}_{{\bf q}=0}-\delta E^{(3)}_{{\bf q}=\sqrt{3}\times\sqrt{3}}$ and, in particular, the 
transition point between the states in both models in Fig.~\ref{fig:dE3} are very similar, there is a quantitative difference.
As we have noted  previously, the local real-space structure of the degeneracy-lifting terms is different in
the $XXZ$ and  single-ion models, implying a higher-order real-space path needed for generating
the ground-state selection in the latter model.\cite{Zh_RSunpub} 
Fig.~\ref{fig:dE3} provides a strong support to this point, 
as the energy difference is smaller for the single-ion model by a factor $2-8$ for the range of 
$0.3\!<\!(1-\Delta) [D/J]\!<\!1$,  in a qualitative agreement with the  parameter of 
the real-space expansion being $\sim 1/z$ ($z=4$, coordination number). \cite{ChZh14}

\begin{figure}[t]
\centerline{
\includegraphics[width=0.99\columnwidth]{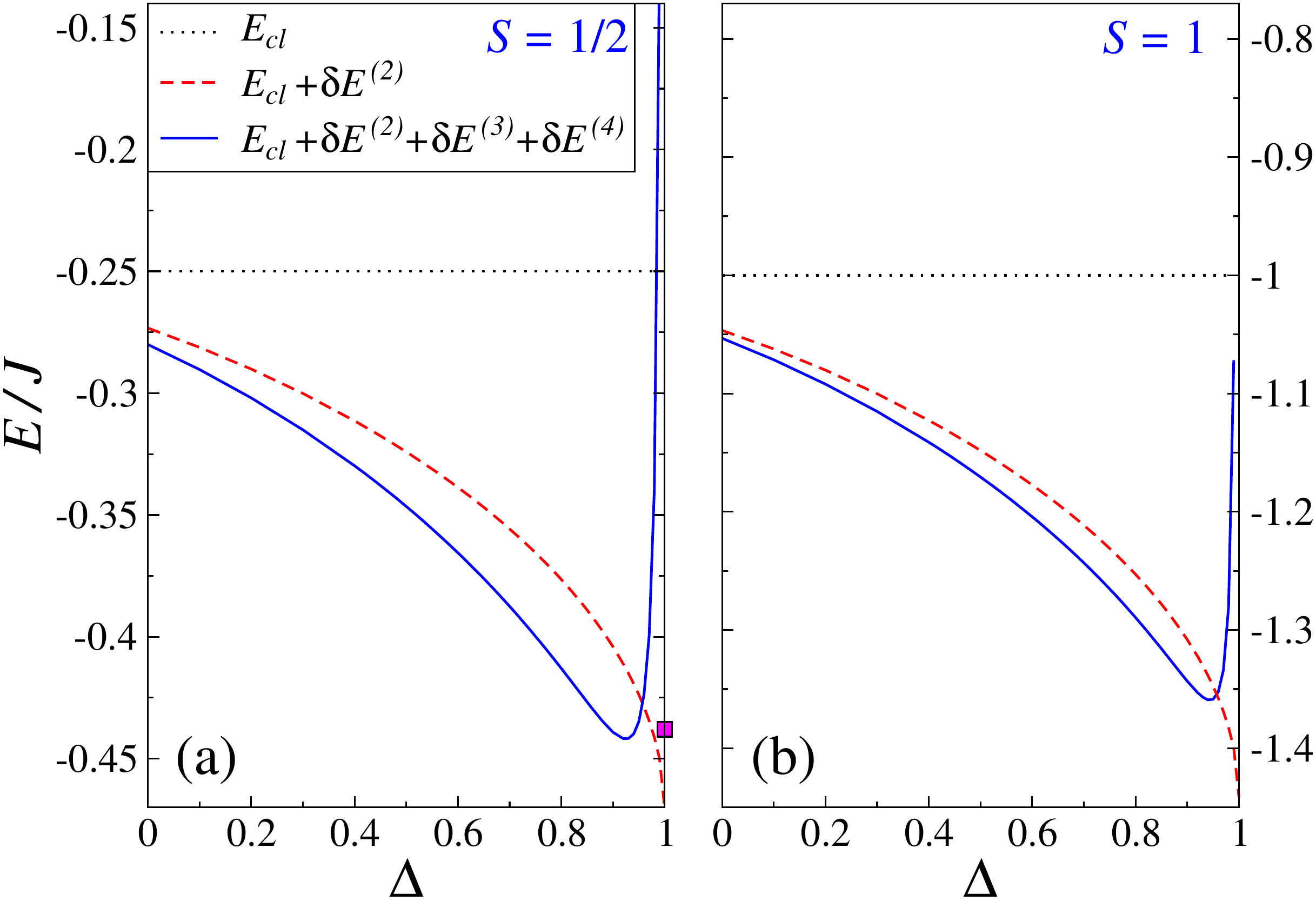}
}
\caption{(Color online)  Energy per site of the ${\bf q}\!=\!0$ state in the $XXZ$ model. 
Dotted line is for the classical term $E_{\rm cl}=-JS^2$, dashed is $E_{\rm cl}+\delta E^{(2)}$ in  (\ref{E2}), and 
solid line is for 
(\ref{E}) with $\delta E^{(3)}$ and $\delta E^{(4)}$ from (\ref{E4}) and  (\ref{dE3s}). Square is the DMRG 
energy for the spin-liquid state.\cite{Yan11} (a) $S=1/2$, (b) $S=1$.
}
\label{fig:dEtot}
\end{figure}

We conclude the discussion of the ground state with Fig.~\ref{fig:dEtot}, which shows 
the energy of the ${\bf q}\!=\!0$ state in (\ref{E}) vs $\Delta$ for the $XXZ$ model for 
two representative spin values $S\!=\!1/2$ and $S\!=\!1$. 
Classical and harmonic contributions to the ground-state energy are also indicated by dotted and dashed lines. 
Note that the energy difference shown in Fig.~\ref{fig:dE3} would 
be nearly invisible on the scale of the plot in Fig.~\ref{fig:dEtot}. We also note that although 
$\delta E^{(3)}$ and $\delta E^{(4)}$ from (\ref{E4}) and (\ref{dE3s}) represent the entire contribution of the ${\cal O}(1)$
order in the $1/S$ expansion of the ground-state energy (\ref{E}), 
their divergences at $\Delta\rightarrow 1$ do not cancel completely, thus signifying
a breakdown of the standard $1/S$ expansion at the Heisenberg limit because of the vanishing 
energy of the flat mode.\cite{Chubukov92}

\subsection{Spectrum and decays}
\label{Sec:spectrum}

We now turn to the spectral properties of the kagom\'{e}-lattice antiferromagnets.
The goal of our consideration is twofold. First, we would like to highlight an unusual 
spectral property that has to be present in a wide variety of  frustrated spin systems with excitations
featuring flat branches.\cite{Ch15}  Second, we give a detailed account of such spectral 
properties in a specific model that describes  Fe-jarosite.\cite{Matan06,Yildirim06}
Because of that we concentrate on the out-of-plane DM-anisotropy model, (\ref{Hs}) with (\ref{HDM}), 
of the $S\!=\!5/2$ nearest-neighbor  kagom\'e-lattice antiferromagnet, which 
closely describes Fe-jarosite in the range of small $D_z$. 
One can expect the results of this consideration to be similar to the ones for 
the $XXZ$ and single-ion anisotropy models given the similarity of their harmonic spectra and anharmonic terms, 
even though the ground-state selection is more subtle in the later models.

\subsubsection{Formalism and a qualitative discussion}
\label{Sec:spectrum_formalism}

Regardless of the model,  using standard diagrammatic rules with the cubic terms in
(\ref{Hss}) and (\ref{Hd}) produces  the spin-wave self-energy  in the form 
\begin{eqnarray}
\Sigma_{\mu,{\bf k}}(\omega)&=&\frac{1}{2N}\sum_{{\bf q},\nu\eta}
\Bigg(\frac{|\Phi^{\nu\eta\mu}_{{\bf q},{\bf k}-{\bf q};{\bf k}}|^2}
{\omega - \varepsilon_{\nu,\bf q} - \varepsilon_{\eta,{\bf k}-{\bf q}}+i\delta}
\label{Sigma}
\\
&&\phantom{\frac{1}{2}\sum_{{\bf q},\nu\eta}\Bigg(}
-\frac{|V^{\nu\eta\mu}_{{\bf q},-{\bf k}-{\bf q},{\bf k}}|^2}
{\omega + \varepsilon_{\nu,\bf q} + \varepsilon_{\eta,{\bf k}-{\bf q}}-i\delta}\Bigg)
 ,\nonumber
\end{eqnarray}
where the first and the second terms are the decay and the source self-energies. 
Because of the summation over the magnon branches $\nu$ and  $\eta$ 
in the decay and source loops, there are nine terms 
in the sum in (\ref{Sigma}), only six of which are distinct. Note that for the DM model one has to 
change $J\!\rightarrow\!J\!+\!D_z/\sqrt{3}$ in the vertices.

Taken on-shell, $\omega\!=\!\varepsilon_{\mu,{\bf k}}$,  the self-energy  (\ref{Sigma})
represents a strictly $1/S$ correction to the magnon energy from the cubic terms, 
$\Sigma_{\mu,{\bf k}}(\varepsilon_{\mu,{\bf k}})\!=\!O(S^0)$. 
The other term of the same order in the  $1/S$-expansion is from the quartic terms, $\varepsilon^{(4)}_{\nu,\bf k}$, 
which was obtained for the DM model in (\ref{EHF4}). 
 Then, the magnon Green's function for the branch $\mu$ can be written as 
\begin{eqnarray}
G^{-1}_\mu({\bf k},\omega)=\omega-\varepsilon_{\mu,{\bf k}}-\varepsilon^{(4)}_{\mu,\bf k} -
\Sigma_{\mu,{\bf k}}(\omega)\,,
\label{GF}
\end{eqnarray}
which, generally, allows to evaluate the spectral function 
$A_\mu({\bf k},\omega)=-(1/\pi){\rm Im}G_\mu({\bf k},\omega)$ 
of the corresponding spin-wave branch $\mu$. 

Since we are interested in the large-$S$ limit and in the resonance-like decay
phenomenon in the spectrum, the following simplification can be used.
Given the off-resonance character of the source term, the frequency-independence of the quartic terms, 
and the large-$S$ limit of the problem, one can neglect the real part of the $1/S$ corrections 
to the spectrum as a first step,  and approximate the self-energy in (\ref{Sigma}) by its 
on-shell imaginary part, i.e. $\Sigma_\mu({\bf k},\omega)\!\approx\! i {\rm Im}\Sigma_\mu({\bf k},\varepsilon_{\mu,{\bf k}})
\! =\! -i\Gamma_{\mu,{\bf k}}$,
with
\begin{eqnarray}
\Gamma_{\mu,{\bf k}}=\frac{\pi}{2N}\sum_{{\bf q},\nu\eta}
|\Phi^{\nu\eta\mu}_{{\bf q},{\bf k}-{\bf q};{\bf k}}|^2
\delta\left(\varepsilon_{\mu,\bf k} - \varepsilon_{\nu,\bf q} - \varepsilon_{\eta,{\bf k}-{\bf q}}\right),
\label{Gamma}
\end{eqnarray}
where  the summation is over magnon branches of the decay products. 
Using the dimensionless vertices in (\ref{V3d1}) and frequencies in (\ref{w1DM}) and (\ref{w23DM}), 
one can rewrite (\ref{Gamma}) as
\begin{eqnarray}
\frac{\Gamma_{\mu,{\bf k}}}{J}=\frac{3\pi}{8N} \sum_{{\bf q},\nu\eta}
|\widetilde{\Phi}^{\nu\eta\mu}_{{\bf q},{\bf k}-{\bf q};{\bf k}}|^2
\delta\left(\omega_{\mu,\bf k} - \omega_{\nu,\bf q} - \omega_{\eta,{\bf k}-{\bf q}}\right),
\label{Gamma1}
\end{eqnarray}
which is explicitly independent of the spin value $S$. As is mentioned above, 
the summation over magnon branches in (\ref{Gamma1})  contains six distinct terms, or potential 
``decay channels.'' A particular decay channel may or may not be  contributing to the decays, depending 
on the kinematic conditions discussed below.  

Before we divert our attention to a specific model, the following qualitative consideration can be made.
The form in (\ref{Gamma1}) suggests that $\Gamma_{\mu,{\bf k}}={\cal O}(J)$, which
is small compared to $\varepsilon_{\mu,\bf k}={\cal O}(JS)$ in the large-$S$ limit.
Thus, generally, one can expect a regular, perturbative $1/S$-effect 
of broadening of the higher-energy part of the dispersive branches due to decays 
into the lower-energy states.\cite{RMP13} However, a  spectacular exception to this rule must occur if 
both of the decay products, branches $\nu$ and  $\eta$,   belong to   flat modes with $\omega_1=const$.
Clearly,  this decay channel produces an essential singularity in  $\Gamma_{\mu,{\bf k}}$ and 
$\Sigma_\mu({\bf k},\omega)$ at the energy equal to   twice   the flat mode energy, the effect that can be seen as a 
resonance-like decay. Therefore, a special care must be exercised in this case as any residual dispersion of the 
flat mode becomes crucial in regularizing such a singularity.

In fact, the way of removing this  singularity is offered by the same $1/S$ fluctuations due to cubic terms,
as the   self-energy (\ref{Sigma}) also yields a dispersion for  the flat mode. This can also be
interpreted as a fluctuation-generated further-neighbor $J_2$ spin-spin interactions,\cite{Chubukov92,ChZh14} 
and such a dispersion of the flat mode has been observed in Fe-jarosite.\cite{Matan06}
Still, this scenario implies that the fluctuation-induced flat-mode bandwidth, $W_1\propto{\cal O}(1)$, 
is of the order $1/S$ compared to the bandwidths of the normal dispersive modes, $W_{2(3)}\propto{\cal O}(S)$. 

Introducing this $1/S$ dispersion for the flat modes in  (\ref{Gamma1}),  suggests that 
 the near-resonance decay-induced broadening of the 
dispersive mode within the energy window of the width $2W_1$ near $2\omega_1$
 should now scale as $\Gamma_{\mu,{\bf k}}={\cal O}(JS)$. This is the \emph{same} energy scale 
 as the spin-wave dispersion itself without any obvious additional smallness. 
Therefore, our analysis implies a very strong broadening,
likely  eliminating spectral weight nearly completely from the respective 
energy range even when spin $S$ is large, providing an example of a spectacular quantum effect in 
an almost classical system.

Thus, in theory, as $S\rightarrow\infty$, we predict that an anomalous broadening and a wipe-out of the spectral 
weight should remain in the spectra of the flat-band frustrated systems, albeit in the energy window of
order ${\cal O}(J)$ while the spectrum width  grows as ${\cal O}(JS)$.
In practice, we argue that such a spectacularly strong quantum effect, leading 
to the quasiparticle breakdown with characteristic termination points and  
ranges of energies dominated by broad continua, must be present in an  almost classical $S\!=\!5/2$ 
kagom\'e-lattice  Fe-jarosite.

\subsubsection{Decay channels in Fe-jarosite}
\label{Sec:decay_conditions}

An extensive analysis of the  kinematic decay conditions in  representative models of the frustrated spin systems
has been provided previously.\cite{RMP13,triPRB09}
Here, the modification of the problem is in having three different magnon branches, which modifies these
condition to
\begin{eqnarray}
\omega_{\mu,\bf k} =\omega_{\nu,\bf q} + \omega_{\eta,{\bf k}-{\bf q}}\, ,
\label{kinematic}
\end{eqnarray}
so that  every decay channel $\mu\!\rightarrow\!\{\nu,\eta\}$ has to be analyzed separately. 
That is, for each branch $\mu$ up to six  channels can be contributing 
independently to the decay rate
\begin{eqnarray}
\Gamma_{\mu,{\bf k}}=\sum_{\{\nu,\eta\}} \Gamma_{{\bf k},\mu\rightarrow \{\nu,\eta\}}\, .
\label{Gamma2}
\end{eqnarray}

\begin{figure}[t]
\includegraphics[width=0.9\columnwidth]{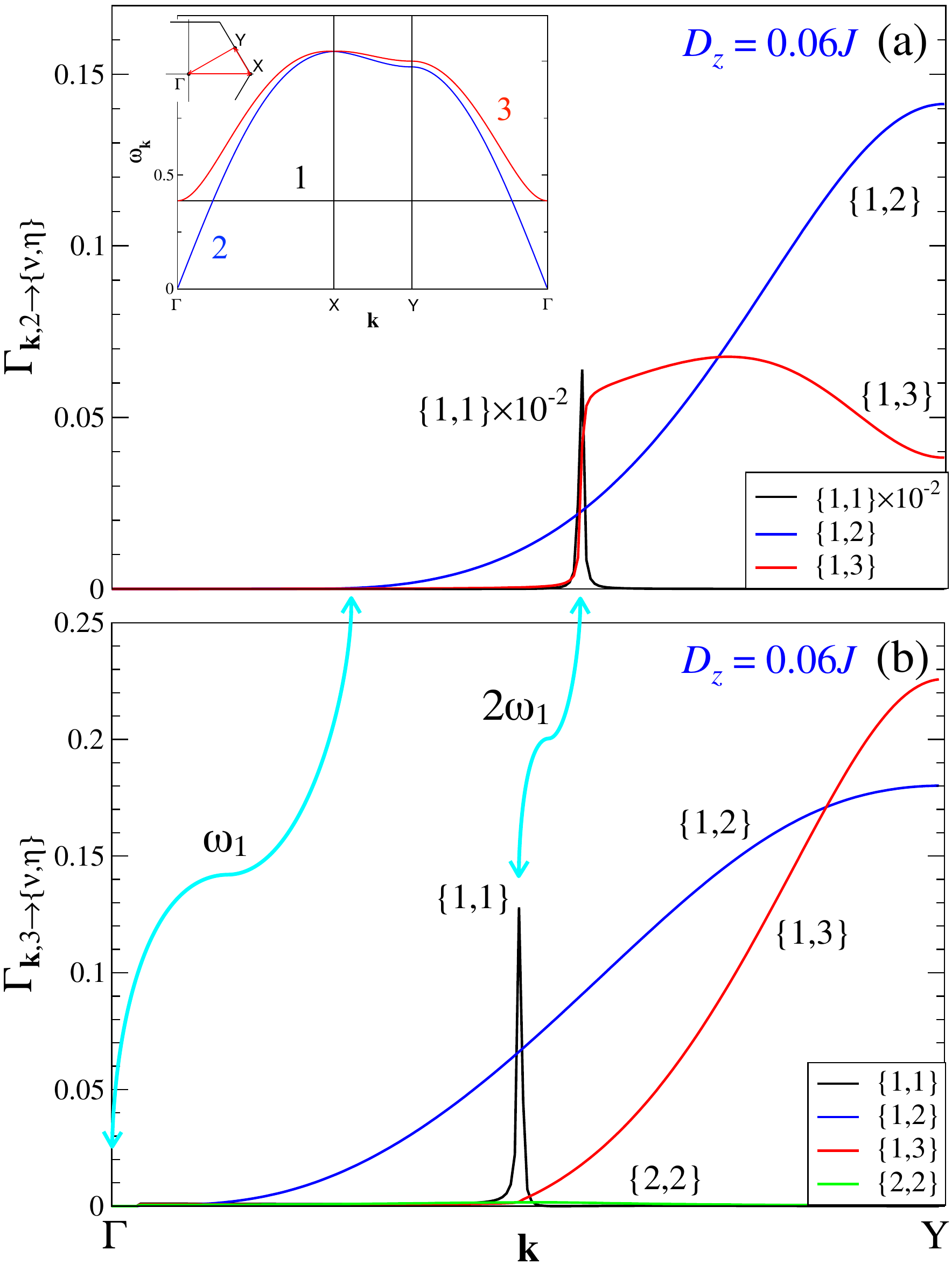}
\caption{(Color online) Individual contributions to the decay rate in (\ref{Gamma1}) from the 
decay channels   $\Gamma_{{\bf k},\mu\rightarrow \{\nu,\eta\}}$ in units of $J$, 
(a) for mode 2, (b) for mode 3 in for the DM model. $D_z\!=\!0.06J$, ${\bf k}$ is along the $\Gamma$Y direction,
and artificial broadening $\delta\!=\!0.002$ has been used. 
${\bf k}$-values that correspond to $\omega_{1}$ and $2\omega_{1}$ are indicated by the arrows. 
Results for the ``resonant'' decay
channel $2\!\rightarrow\!\{1,1\}$ are scaled down by the factor  $10^{-2}$.  
Inset: harmonic frequencies  
$\omega_{\alpha,\bf k}$ along the representative cuts of the BZ for all three modes from 
Fig.~\ref{Fig:wk_J2}.
}
\label{Fig:GkGY}
\end{figure}

While the decay conditions are model-specific, Fe-jarosite offers a commonplace scenario: a
predominantly nearest-neighbor Heisenberg system with  subleading symmetry-breaking anisotropic term,
which is responsible for lifting the flat mode to a finite energy. We, therefore, analyze the decay channels 
for this representative situation for a small $D_z=0.06J$, used in fitting  spin-wave
spectrum of Fe-jarosite.\cite{Yildirim06}
The harmonic dispersions for this value of $D_z$ are shown in Fig.~\ref{Fig:wk_J2} and in the inset of 
Fig.~\ref{Fig:GkGY}(a), and we use them in our analysis. 

We  use the same numeration of the branches as before:  1=flat mode,  2=gapless mode (Goldstone branch), 
and 3=gapped dispersing mode.
While the flat mode does have one active decay channel into two Goldstone modes, $1\!\rightarrow\!\{2,2\}$, 
for a small range of momenta near the center of the Brillouin zone, 
${\bf k}\!<\!{\bf k}_c$ where ${\bf k}_c$ is the intersect point of the branches, see Fig.~\ref{Fig:GkGY}(a), 
its effect is truly minor and the main interest is in the decays of the dispersive modes.
From the picture of harmonic energies, it is easy to
see that the energy conservation in (\ref{kinematic}) can be satisfied for three (four)
decay channels out of the six for the mode 2(3), one being the ``resonance-like,'' 
$2(3)\!\rightarrow\!\{1,1\}$, mentioned above, and two (three) are ``regular.'' The latter are 
$2(3)\!\rightarrow\!\{1,2\}$ and $2(3)\!\rightarrow\!\{1,3\}$, with an additional 
channel for the mode $3\!\rightarrow\!\{2,2\}$. 

Using  (\ref{Gamma1}), we calculate 
contributions of the individual decay channels,  $\Gamma_{{\bf k},\mu\rightarrow\{\nu,\eta\}}$, 
shown in Figs.~\ref{Fig:GkGY}(a) and (b), 
for   modes 2 and 3 and for a representative ${\bf k}$-cut of the Brillouin zone.
One can see that for $2(3)\!\rightarrow\!\{1,1\}$ it gives a $\delta$-peak at $2\omega_{1}$, that 
$2(3)\!\rightarrow\!\{1,2\}$ channel has a threshold at $\omega_{1}$ and 
$2(3)\!\rightarrow\!\{1,3\}$ at $2\omega_{1}$. 
In Fig.~\ref{Fig:GkGY}(a), the resonance decay rate is scaled down by $10^{-2}$ and would otherwise 
dwarf the rest even for the
small  artificial broadening of the $\delta$-function in (\ref{Gamma1}).

A somewhat unexpected finding  is a strong suppression of the $3\!\rightarrow\!\{2,2\}$  
and the $3\!\rightarrow\!\{1,1\}$ channels.
A closer analysis have identified different origins for them. For the $3\!\rightarrow\!\{2,2\}$ channel,
decay products involve  long-wavelength excitations from the Goldstone branch, which provides an extra smallness  
in the decay amplitude compared with the ``regular'' ones. 
For the $3\!\rightarrow\!\{1,1\}$ channel the situation is more subtle.
One can demonstrate that the vertex $\widetilde{\Phi}^{113}_{\bf q,k-q;k}$ carries an extra $D_z$ 
compared with the vertex for the $2\!\rightarrow\!\{1,1\}$ channel, 
yielding a factor of about $1/100$ in the decay rate. 
The physical reason for this suppression can be hypothesized. 
Quasiclassically, the mode 2 is the ``in-plane'' mode, while  modes 3 and  1 
are ``out-of-plane.'' Then the in-plane mode can  couple naturally to the two 
out-of-plane modes, while the out-of-plane mode  has hard time splitting into two. 

Thus, the suppression of the resonant decays of the mode 3   seems to be due to 
a subtle cancellation in the structure of the corresponding vertex. 
We note that for a  realistic case of  Fe-jarosite, other symmetry-breaking terms
are also present, so one can expect  the subtle cancellations of more 
symmetric models to be violated.

\begin{figure}[t]
\includegraphics[width=0.99\columnwidth]{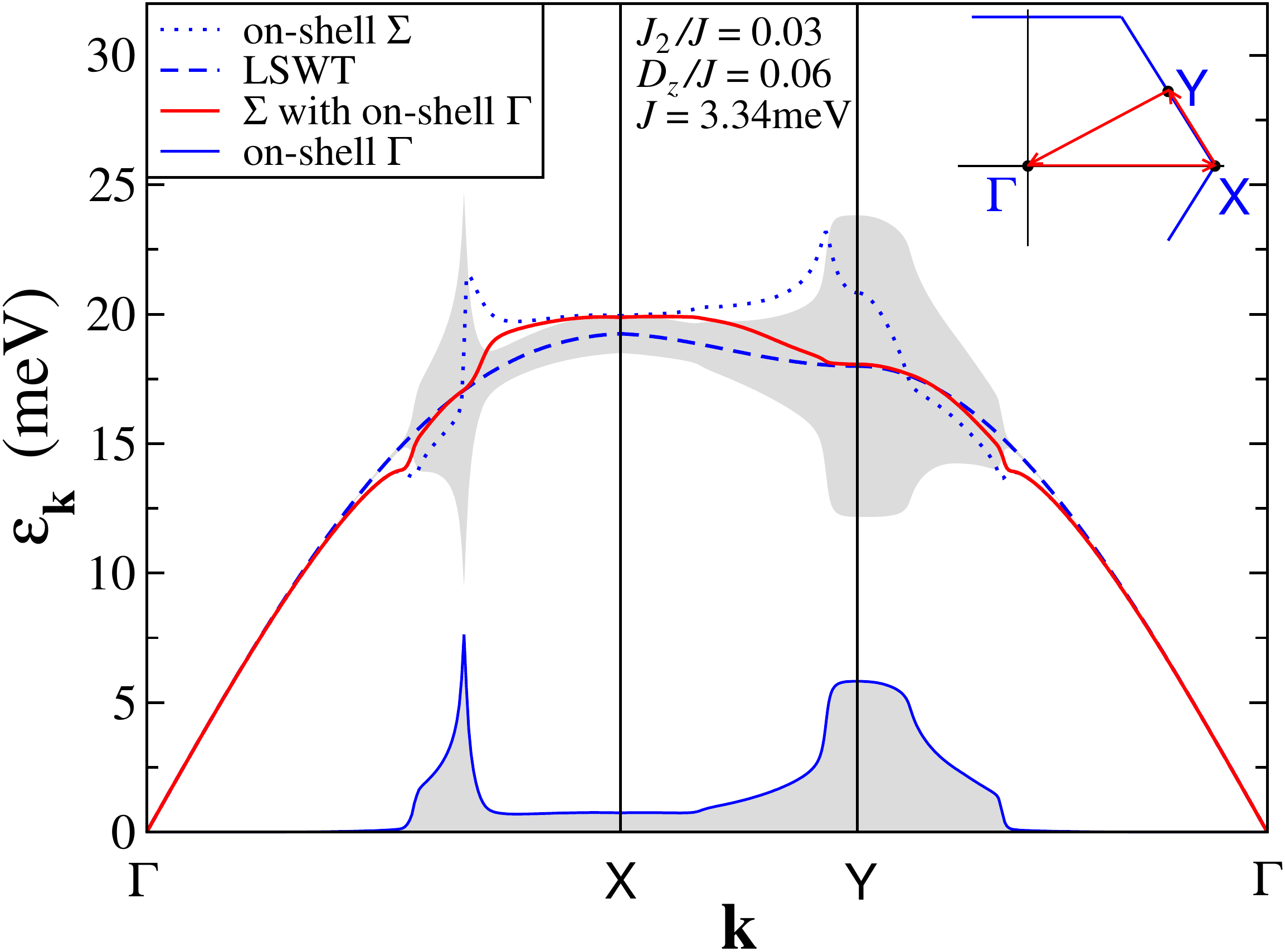}
\caption{(Color online) Lower curve with the shading is the on-shell $\Gamma_{2,{\bf k}}$ from (\ref{Gamma1}).
Dashed line is the linear spin-wave theory energy of the gapless dispersive mode,  $\varepsilon_{2,\bf k}$, from
(\ref{w23DM}). Shaded area shows the  half-width boundaries of a lorentzian peak,
$\varepsilon_{2,\bf k}\pm\Gamma_{2,{\bf k}}$.
Dotted and solid lines are different approximations for the renormalization 
of the real part of the self-energy in (\ref{Sigma}), see text. Parameters are as shown in the plot.
}
\label{Fig:wkFeJ}
\end{figure}

\subsubsection{Spectrum of Fe-jarosite}
\label{Sec:spectrum_FeJ}

We now finally approach the realistic description of the Fe-jarosite spectrum. 
We note that more details of this discussion is offered elsewhere.\cite{Ch15}
As was mentioned above,
the essential singularity in the dispersive modes is naturally removed by the residual dispersion of the 
flat mode.  The experimentally observed dispersion of the flat mode\cite{Matan06} has been fit\cite{Yildirim06} by 
introducing small next-nearest neighbor exchange $J_2=0.03J$, ignoring its possible quantum origin. 
Since we do not perform a fully self-consistent calculation here,  the same approach suffices for the removal of 
the singularity in the decay rate (\ref{Gamma1}). We, thus, modify the flat-mode dispersion according to (\ref{w1J1J2Dz}) 
and ignore other corrections from the $J_2$ term.

In Sec.~\ref{Sec:structure_factor} we elaborate on the possible way of separating contributions of different modes 
in the neutron-scattering structure factor, which allows us to concentrate on individual modes.
Therefore, our Fig.~\ref{Fig:wkFeJ} summarizes the effects of   decays on the gapless dispersive mode only, as they are
most pronounced in it. 
We have used parameters shown in the figure with the value of $J=3.34$meV from the previous work.\cite{Yildirim06}

The lower curve shows the  on-shell $\Gamma_{2,{\bf k}}$ obtained from (\ref{Gamma1})
with the flat-mode dispersion from (\ref{w1J1J2Dz}), and it includes 
all three decay channels discussed above. The dashed line is the 
linear spin-wave energy,  $\varepsilon_{2,\bf k}$, with the shaded area around it representing 
$\varepsilon_{2,\bf k}\pm\Gamma_{2,{\bf k}}$, i.e., half-width at the half-maximum boundaries of a lorentzian peak. 
We have also taken into account renormalization of the real part 
of the self-energy in (\ref{Sigma}), with the dotted line showing the $1/S$ on-shell result and   the solid line including
 effect of self-consistency by   taking into account imaginary part
from $\Gamma_{2,{\bf k}}$ in calculation of Re$\Sigma$. One can see that the resultant 
effects of the nonlinear terms on the real part of the spectrum are relatively minor, 
in agreement with the discussion after Eq.~(\ref{GF}).

\begin{figure}[t]
\includegraphics[width=0.99\columnwidth]{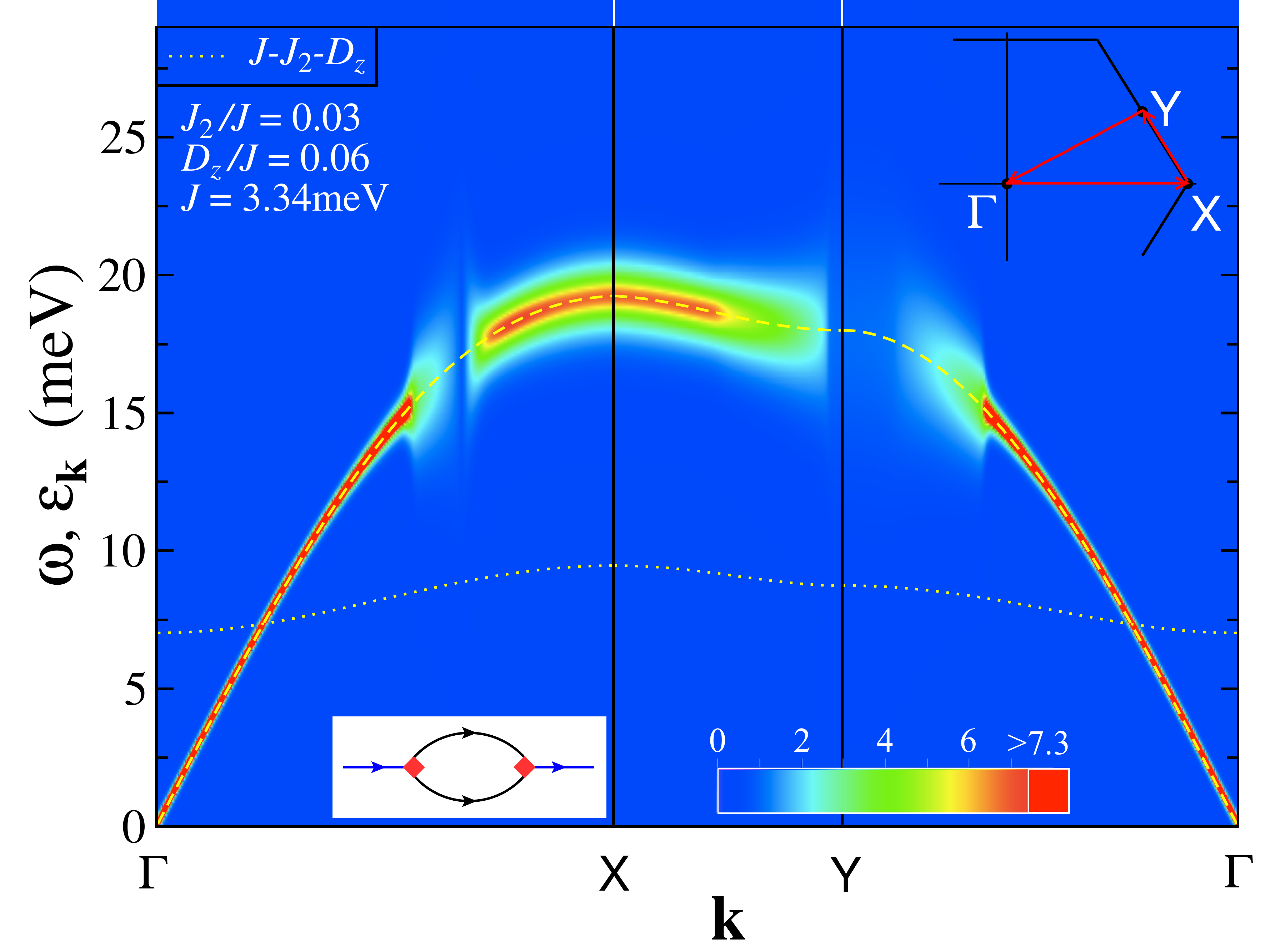}
\caption{(Color online)\ \ 
Intensity plot of  the spectral function $A_2({\bf k},\omega)$ of the gapless mode in units of $(2SJ)^{-1}$
along the same path and for the same parameters as in Fig.~\ref{Fig:wkFeJ},  lines are  from the same figure, 
see text.  A small broadening $\delta/2SJ=0.002$ 
has been added to $\Gamma_{2,{\bf k}}$ in (\ref{Gamma}). 
Left panel inset: decay diagram.
}
\label{fig:Akw}
\end{figure}

We complement Fig.~\ref{Fig:wkFeJ}  by an intensity map of the spectral function, $A_2({\bf k},\omega)$,
for  the same dispersive mode along the same representative path in the Brillouin zone and for the same 
parameters, see Fig.~\ref{fig:Akw}.
Dashed and dotted lines are the spin-wave results from Fig.~\ref{Fig:wkFeJ} and are guides to the eye.
The magnon self-energy is approximated by its on-shell imaginary part, as discussed above.
The upper cut-off of the intensity of the spectral function corresponds to the maximal height of the peaks 
in the non-resonant decay region in Fig.~\ref{Fig:wkFeJ} and  
translates into the broadening $\Gamma_{{\bf k}}\!\approx\!0.73$meV  for the Fe-jarosite, 
easily resolvable by the modern neutron-scattering experiments. 

Given the large spin value, $S=5/2$, we estimate that the ordered moment in  Fe-jarosite 
should be nearly 90\% of its classical value, see Fig.~\ref{deltaS}. It is then very natural to assume that 
the spectral properties should be fully describable by the harmonic spin-wave picture, the point 
of view taken in Ref.~\onlinecite{Yildirim06}.  
Indeed, our Figs.~\ref{Fig:wkFeJ} and \ref{fig:Akw} demonstrate that the spin-wave excitation is 
sharp below the flat-band energy $\varepsilon_{1,{\bf k}}$ 
and  acquires only an infinitesimal  width  for the energies above it. 
However, there is a sharp threshold at twice the bottom of the flat band minimum, 
$2\varepsilon^{\rm min}_{1,{\bf k}}$,  
above which there is a very strong broadening, reaching about one third of the bandwidth,
signifying an overdamped spectrum.\cite{Ch15}
Above $2\varepsilon^{\rm min}_{1,{\bf k}}$ in Fig.~\ref{fig:Akw} there is a broad energy band with the features that 
look like a rip  in the spectrum, consistent with the missing spectral weight  in the experimental data.
This threshold singularity is also remarkably similar to the spectral 
signatures of the quasiparticle breakdown phenomenon 
in quantum Bose liquids and $S=1/2$ spin-liquids. \cite{Zaliznyak,Zheludev}

Although  at the energies above  twice the top of the flat-band maximum, $2\varepsilon^{\rm max}_{1,{\bf k}}$,
the decays are ``regular,'' i.e., occurring due to other, non-resonant channels,  
 they are still providing a measurable broadening to the spectrum, so it is instructive to compare the two regions.
The ``regular'' decays result in the maximal values of $\Gamma$ of order $0.2-0.3J$, an agreement with the similar 
effects in triangular-lattice antiferromagnets\cite{triPRB09} and other frustrated spin systems.\cite{RMP13}
The maximal broadening in the resonant-decay region is $\Gamma \approx 1.7J$ for the Fe-jarosite model,
which is larger than the effect of the ``regular'' decays by a factor close to 5 ($=2S$). This is in a remarkably 
close accord with the qualitative argument on the scaling of the resonance decay rate with the spin value, 
provided after Eq.~(\ref{Gamma1}) above.

\begin{figure}[t]
\includegraphics[width=0.90\columnwidth]{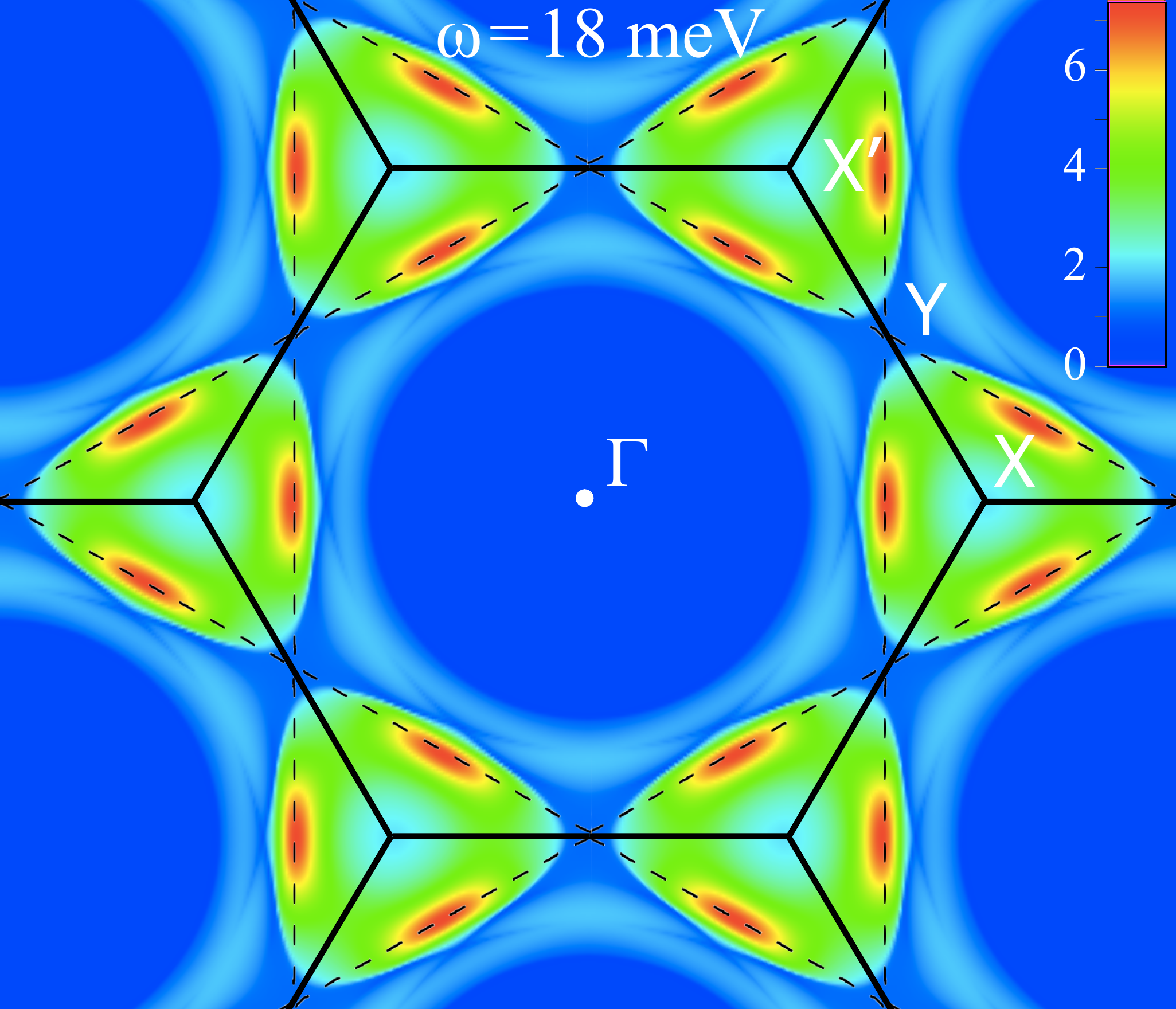}
\caption{(Color online)\ \ 
Intensity map of $A_2({\bf k},\omega)$ in units of $(2SJ)^{-1}$ vs ${\bf k}$ 
throughout the Brillouin zone for for the same set of parameters as in 
Fig.~\ref{Fig:wkFeJ} for $\omega=18$meV. 
The upper cut-off of the intensity scale
corresponds to the broadening $\Gamma_{2,{\bf k}}$   of 0.73~meV  for the Fe-jarosite values of $S$ and $J$.
Dashed black lines are peak positions  from the linear spin-wave theory. 
}
\label{fig:2DoS}
\end{figure}

While a more detailed description of the spectral features of Fe-jarosite is given elsewhere,\cite{Ch15}
we would like to highlight here a different,  perhaps more dramatic view on the drastic transformations 
in its spectrum in the range of energies that falls within the resonant-decay band. In Fig.~\ref{fig:2DoS}
we show an intensity plot of the 
constant-energy cut of the spectral function, $A_2({\bf k},\omega)$,
of   the same dispersive  mode as in Figs.~\ref{Fig:wkFeJ} and \ref{fig:Akw} for the energy 18meV,  
which is close to the middle of the resonant-decay region.
The dashed lines show the expectations from the linear spin-wave theory of where the contours of sharp, 
well-defined peaks should have occurred. Instead, one can observe strong deviation from such expectations, 
characterized by a broadening, massive redistribution of the spectral weight into   
 different regions of ${\bf k}$-space, together with the multitude of intriguing  features, which are 
related to the van Hove singularities in the two-particle density of states of the flat-band decay products.\cite{triPRB09} 
Altogether, Figs.~\ref{Fig:wkFeJ}, \ref{fig:Akw}, and \ref{fig:2DoS} offer a convincing evidence of the highly  
non-trivial and very strong quantum effects in the the dynamical response of a nearly classical flat-band 
kagom\'{e}-lattice antiferromagnet, which are facilitated by the nonlinear couplings.

\subsection{Dynamical structure factor}
\label{Sec:structure_factor}

Inelastic neutron scattering cross-section is directly related to the diagonal components 
of the dynamical structure factor, or the spin-spin dynamical correlation function, given by
\begin{eqnarray}
{\cal S}^{i_0i_0}({\bf q},\omega) = \int_{-\infty}^{\infty} \frac{dt}{2\pi}\,e^{i\omega t}\langle 
S^{i_0}_{\bf q}(t)S^{i_0}_{-\bf q}\rangle  \,,
\label{Sqws}
\end{eqnarray}
where $i_0$ refers to the laboratory frame $\{x_0,y_0,z_0\}$
and 
\begin{eqnarray}
S^{i_0}_{\bf q}=\sum_\alpha S^{i_0}_{\alpha,{\bf q}},
\label{Sq}
\end{eqnarray}
involves summation over the spins $\alpha$ in the unit cell.

Because of the coplanar spin configuration in the considered kagom\'{e}-lattice antiferromagnets, 
transformation from the laboratory reference frame of  $i_0$ to the local spin basis of (\ref{Hs})
yields a mix of different diagonal and off-diagonal terms in the structure factor,\cite{Mourigal13} which  
conveniently separate into the in-plane and out-of-plane parts of ${\cal S}^{\rm tot}({\bf q},\omega)$. 
Assuming equal contribution of all three $i_0$  components to the neutron-scattering cross section\cite{Mourigal13}
and using the  mapping of spins on bosons (\ref{HP}) allows one to perform a straightforward $1/S$ ranking
of different contributions to the structure factor, in which the transverse components are, as usual, dominate 
the longitudinal and mixed terms. 

\begin{figure}[tb]
\includegraphics[width=0.99\columnwidth]{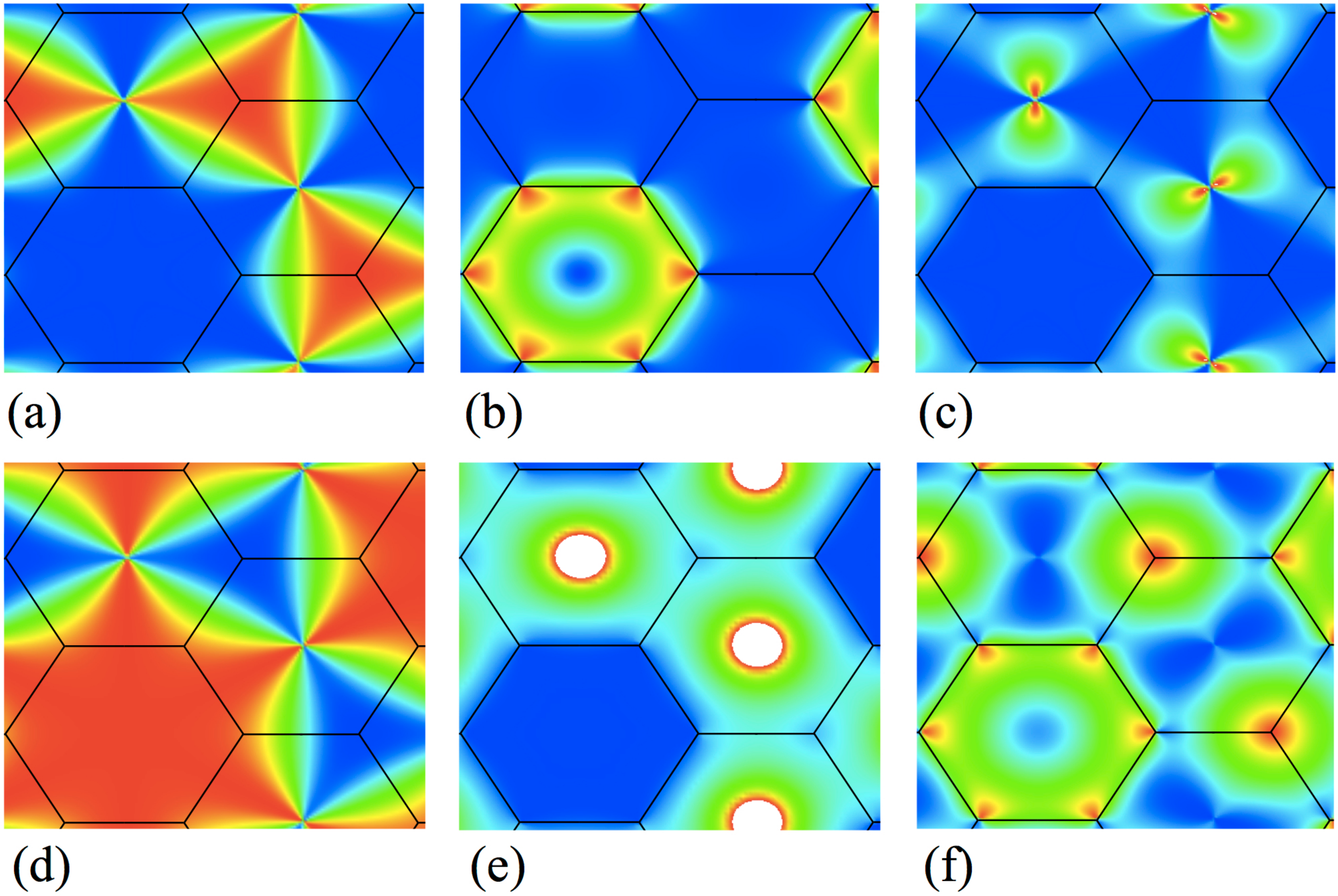}

\caption{(Color online) Kinematic formfactors: 
$F^{\rm out}_{\nu\bf q}$
for the (a) flat, (b) dispersive gapless, (c) dispersive gapped mode;
$F^{\rm in}_{\nu\bf q}$
for the (d) flat, (e) dispersive gapless, (f) dispersive gapped mode. $F^{\rm in}_{2\bf q}$ in (e) diverges at 
some $\Gamma$ points as $\omega^{-1}_{2,{\bf q}}$.
}
\label{Fig:Fnuqout}
\end{figure}

The subsequent algebra involves the two-step transformation (\ref{linearT}) and (\ref{BogolyubovT}) from 
the Holstein-Primakoff bosons to the quasiparticles,  yielding the leading contributions to the structure factor as
directly related to the spin-wave spectral functions $A_\nu({\bf q},\omega)$
\begin{eqnarray}
{\cal S}^{\rm in(out)}({\bf q},\omega) \approx\sum_{\nu} F^{\rm in(out)}_{\nu\bf q} A_\nu({\bf q},\omega) \,,
\label{Sqws1}
\end{eqnarray}
where we introduced  kinematic formfactors 
\begin{eqnarray}
&&F^{\rm in}_{\nu\bf q} = \frac{S}{2} (u_{\nu\bf q}+v_{\nu\bf q})^2
\left(1- R_{\nu{\bf q}}\right)    \,,\nonumber\\
&&F^{\rm out}_{\nu\bf q} = \frac{S}{2} (u_{\nu\bf q}-v_{\nu\bf q})^2
\left(1+2R_{\nu{\bf q}}\right)\,,
\label{Fnuqs}
\end{eqnarray}
with
\begin{eqnarray}
R_{\nu{\bf q}}=\frac12\sum_{\alpha\neq\alpha'}w_{\nu,\alpha}({\bf q})w_{\nu,\alpha'}({\bf q})\,.
\label{Rnuqs}
\end{eqnarray}

\begin{figure}[t]
\includegraphics[width=0.9\columnwidth]{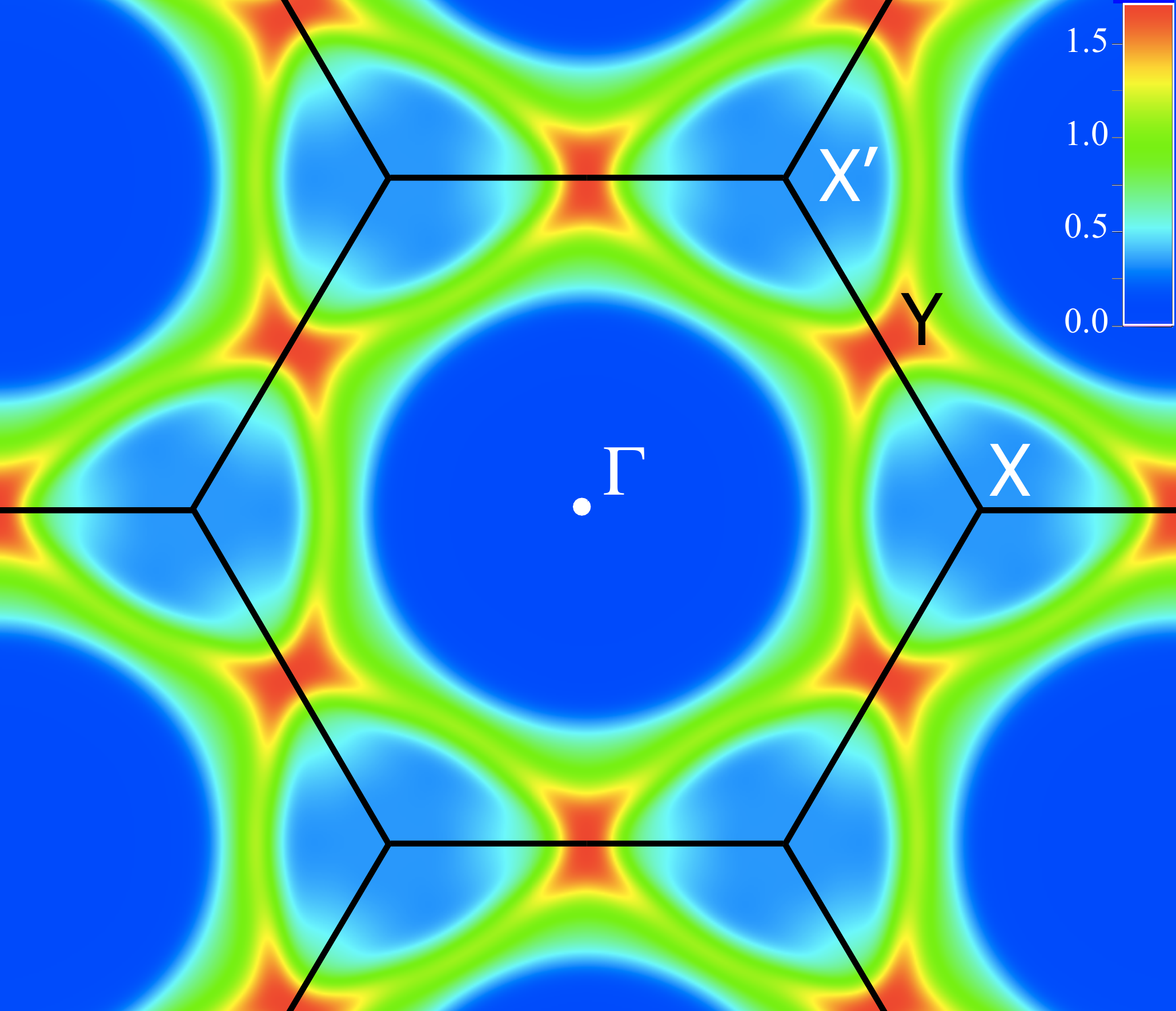}
\caption{(Color online)\ \ 
A 2D map of the inverse lifetime  from (\ref{Gamma1a}) in units of $J$
in the ${\bf q}$-space. A small artificial gaussian 
${\bf q}$-broadening with $\sigma=0.02\pi/a$ was used to mimic instrumental resolution. 
The scale is given in the inset and the maximal value of $\Gamma^{\rm max}_{2,{\bf q}}\approx 1.7J$ corresponds to 
about 6meV for Fe-jarosite. 
}
\label{Fig:GkGY3}
\end{figure}

An important property of the kinematic formfactors in (\ref{Fnuqs}) is their ${\bf q}$-dependence. 
They are modulated in the ${\bf q}$-space  and are typically suppressed 
in one of the Brillouin zones while are maximal in the others. This effect is characteristic to the 
neutron-scattering in the non-Bravais lattices and  is, in a way, similar to the effect 
known as the Bragg peak extinction for the elastic scattering in such lattices. Because of that property, 
one may be able to focus on a specific excitation branch without intermixing contributions from the others 
by selecting a particular component of the structure factor in a particular Brillouin zone.

Our Fig.~\ref{Fig:Fnuqout} shows $F^{\rm in(out)}_{\nu\bf q}$
for   three branches of excitations in the Fe-jarosite model. It demonstrates that such a  ``filtering out'' can be quite useful.
For example, the out-of plane component of the structure factor, ${\cal S}^{\rm out}({\bf q},\omega)$, 
should be dominated  in one of the three distinct 
Brillouin zones by  the spectral function $A_2({\bf q},\omega)$ of 
 only one of the dispersive modes. This feature can be utilized in the neutron-scattering experiments.  

Lastly, we highlight another quantitative way of analyzing  ${\cal S}({\bf q},\omega)$.
Assuming that one can focus on a particular excitation branch as mentioned above,  one can 
suggest that a faithful representation of the 
inverse lifetime (linewidth) $\Gamma_{\mu{\bf q}}$ from (\ref{Gamma}) and its distribution in ${\bf q}$-space across the 
Brillouin zone can be obtained from the moments of the dynamical structure factor as \cite{Broholm}
\begin{eqnarray}
\widetilde{\Gamma}_{{\bf q}}=\sqrt{\langle\omega^2{\cal S}({\bf q},\omega)\rangle-
\langle\omega{\cal S}({\bf q},\omega)\rangle^2},
\label{Gamma1a}
\end{eqnarray}
where $\langle\omega^n{\cal S}({\bf q},\omega)\rangle\!=\!\int  \omega^n{\cal S}({\bf q},\omega)d\omega$
are the moments of the structure factor and  normalization $\langle{\cal S}({\bf q},\omega)\rangle\!=\!1$ is assumed.
Then, this experimentally unbiased procedure would allow extracting a 2D ${\bf q}$-map of the quasiparticle
broadening. We demonstrate the effectiveness of this approach in Fig.~\ref{Fig:GkGY3}, which shows 
an example of such a map derived using the  procedure in (\ref{Gamma1a}) from a \emph{gaussian} form 
$A_2({\bf q},\omega)$, i.e. the gaussian function with a maximum at $\varepsilon_{2,{\bf q}}$ and the width  
$\Gamma_{2,{\bf q}}$. As expected, the extracted map of $\widetilde{\Gamma}_{{\bf q}}$ is nearly
identical to the map of $\Gamma_{2,{\bf q}}$ itself. However, for a more natural lorentzian form of  
$A_2({\bf q},\omega)$ one can show that the extracted map corresponds to 
$\widetilde{\Gamma}_{{\bf q}}\!\propto\!\sqrt{\omega_{\rm max}\Gamma_{2,{\bf q}}}$, where 
$\omega_{\rm max}$ is the upper limit of the integration over $\omega$ in the moments in (\ref{Gamma1a}).
Importantly, this result is still providing a direct information on the quasiparticle broadening map, albeit 
on a different scale. Therefore, aside from demonstrating at which momenta  the decays are most intense,
the suggested procedure  also provides another way of ``fingerprinting'' of the broadening
due to resonance-like decay into the flat modes.

\section{Conclusions}

In summary, we have provided a systematic consideration of the nonlinear  $1/S$ expansion 
of several anisotropic models of the kagom\'{e}-lattice antiferromagnets. 
We have demonstrated the role of the nonlinear terms in the 
quantum order-by-disorder selection of the ground state and 
presented a strong evidence of the rare case of quantum and thermal 
fluctuations favoring  different ground states in two of these models.
We have  provided a detailed analysis of the  excitation spectrum 
 of the   $S=5/2$ iron-jarosite to illustrate our proposed general scenario of 
 drastic transformations in the spectra of the flat-band frustrated magnets. 
Our study calls for further neutron-scattering experiments in these 
systems.

\begin{acknowledgments}

We acknowledge useful  discussions with Collin Broholm,  Christian Batista, Federico Becca, 
Andrey Chubukov,  Andreas L\"auchli, Young Lee, Kittiwit Matan,
George Jackeli, Steven White, and Zhenyue Zhu. 
This work was supported by the U.S. Department of Energy,
Office of Science, Basic Energy Sciences under Award \# DE-FG02-04ER46174.
A.~L.~C. would like to thank Aspen Center for Physics and the Kavli Institute for Theoretical Physics
where different stages of this work were advanced. 
The work at Aspen was supported by NSF Grant No. PHYS-1066293 and 
the research at KITP was supported by NSF Grant No. NSF PHY11-25915.

\end{acknowledgments}

\appendix

\section{Hartree-Fock corrections}
\label{AppA}

The quartic terms in (\ref{Hxxz4}) yield a correction to the
ground-state energy given by the four-boson averages, which are decoupled into the products of the
binary Hartree-Fock averages (\ref{HF}) using  Wick's theorem
\begin{eqnarray}
&& \langle a^\dagger_i a_i a^\dagger_j a_j\rangle = n^2+m^2+\bar\Delta^2\ ,   \\
&& \langle a^\dagger_i a_i a_i a_j\rangle = 2n\bar\Delta +  m\delta, \ \
\langle a^\dagger_j a^\dagger_j a_j a_i\rangle = 2nm + \bar\Delta\delta,\nonumber 
\end{eqnarray}
where the Hartree-Fock averages are obtained similarly to the calculation of the staggered magnetic moment in (\ref{Sav1}).
Namely, for the on-site averages $n=\langle a_{\alpha,\ell}^\dag a^{\phantom{\dag}}_{\alpha,\ell}\rangle$ and
$\delta=\langle a_{\alpha,\ell}^{\phantom{\dag}} a^{\phantom{\dag}}_{\alpha,\ell}\rangle$, following the 
transformations (\ref{linearT}) and (\ref{BogolyubovT}) from $a_\alpha$ to $d_\mu$ and to $b_\mu$
and using equivalence of all three sublattices one arrives to the symmetrized expressions
\begin{eqnarray}
&&n=\langle a_{\alpha,\ell}^\dag a^{\phantom{\dag}}_{\alpha,\ell}\rangle=\frac{1}{3N}
\sum_{\mu,{\bf k}} v^2_{\mu{\bf k}}\,, 
\nonumber\\
&&\delta=\langle a_{\alpha,\ell}^{\phantom{\dag}} a^{\phantom{\dag}}_{\alpha,\ell}\rangle 
=\frac{1}{3N}\sum_{\mu,{\bf k}} u_{\mu{\bf k}}v_{\mu{\bf k}},
\label{AHF1}
\end{eqnarray}
with $u_{\mu{\bf k}}$ and $v_{\mu{\bf k}}$ from (\ref{Bogolyubov_uv}). 
For the nearest-neighbor two-site averages, the same transformations lead to 
\begin{eqnarray}
 &&m = \langle a_{\alpha,\ell}^\dag a^{\phantom{\dag}}_{\beta,\ell'}\rangle=
\frac{1}{N}\sum_{\mu,{\bf k}} f_{\beta\alpha}({\bf k})
v^2_{\mu{\bf k}}\,, \nonumber\\
\label{AHF2}\\
&& \bar\Delta = \langle a_{\alpha,\ell}^{\phantom{\dag}} a^{\phantom{\dag}}_{\beta,\ell'}\rangle 
=\frac{1}{N}\sum_{\mu,{\bf k}} f_{\beta\alpha}({\bf k}) u_{\mu{\bf k}}v_{\mu{\bf k}},\nonumber
\end{eqnarray}
where $f_{\beta\alpha}({\bf k})=\cos(k_{\beta\alpha}) \,
w_{\mu,\alpha}({\bf k}) w_{\mu,\beta}({\bf k})$, 
$\alpha\neq\beta$, $w_{\mu,\alpha}({\bf k})$ are the components of the eigenvector in the transformation 
(\ref{linearT}), and $k_{\beta\alpha}={\bf k}\bm{\rho}_{\beta\alpha}$ with
  $\bm{\rho}_{\beta\alpha} = \bm{\rho}_\beta - \bm{\rho}_\alpha$ as before.
Although  it seems that the two-site averages may depend on the choice of  $\alpha$ and $\beta$,
one can verify that all three possible combinations of  $\alpha\neq\beta$ pairs yield the same answer.

Quartic terms also yield  Hartree-Fock contributions to the linear spin-wave dispersions via 
(\ref{EHF4}). The corresponding constants for the Heisenberg and DM terms in (\ref{H24a}) are
\begin{eqnarray}
&&C_1= -n +  \frac{3\bar\Delta}{2} - \frac{m}{2}, \ \ C_2= -m +  \frac{3\delta}{4} - \frac{n}{2},\nonumber\\
&&C_3= -\bar\Delta +  \frac{3n}{2} - \frac{\delta}{4}, \ \ C_4=  \frac{3m}{8} - \frac{\bar\Delta}{8},
\label{C4}\\
&&D_1= n -  \frac{\bar\Delta}{2} - \frac{m}{2}, \ \ D_2= m -  \frac{\delta}{4} - \frac{n}{2},\nonumber\\
&&D_3= \bar\Delta - \frac{n}{2} - \frac{\delta}{4}, \ \ D_4=  -\frac{m}{8} - \frac{\bar\Delta}{8}. \nonumber
\end{eqnarray}



\begin{thebibliography}{99}
\bibitem{Broholm} We are thankful to Collin Broholm for this suggestion.
\end{thebibliography}
\end{document}